\documentclass[a4paper,fleqn,usenatbib]{mnras}

\usepackage{newtxtext,newtxmath}

\usepackage[T1]{fontenc}
\usepackage{ae,aecompl}

\usepackage{graphics,epsfig,psfig}
\usepackage{dblfloatfix}
\usepackage[normalem]{ulem}
\usepackage[dvipsnames]{color}
\usepackage{xcolor}
\usepackage[]{inputenc,amssymb}
\usepackage{multirow, multicol}
\usepackage{booktabs,caption}
\usepackage{lipsum}
\usepackage{widetext}
\usepackage{pifont}
\usepackage{natbib}
\usepackage{dblfloatfix,afterpage}
\usepackage[normalem]{ulem} 
\usepackage{graphicx}	
\usepackage{amsmath}	
\usepackage{amssymb}	

\title[Acceleration of wind vs SNe ejecta in star clusters]{
Realistic modeling of wind and supernovae shocks in star clusters: addressing ${\rm ^{22}Ne/^{20}Ne}$ and other problems in Galactic cosmic rays}
\author[Gupta et al.]{
Siddhartha Gupta$^{1,2}$\thanks{E-mail:siddhartha.gupta19@gmail.com},
Biman B. Nath$^{1}$\thanks{E-mail:nath.biman@gmail.com},
Prateek Sharma$^{2,3}$,
David Eichler$^{4}$
\\
$^{1}$Raman Research Institute, Sadashiva Nagar, Bangalore 560080 India\\
$^{2}$Department of Physics, Indian Institute of Science, Bangalore  India\\
$^{3}$MPI f\"{u}r Astrophysik, Karl-Schwarzschild str 1, D-85741 Garching, Germany\\
$^{4}$Department of Physics, Ben-Gurion University, Beer Sheva, Israel\\
}

\date{Accepted XXX. Received YYY; in original form ZZZ}

\begin{document}
\label{firstpage}
\pagerange{\pageref{firstpage}--\pageref{lastpage}}
\maketitle

\begin{abstract}
Cosmic ray (CR) sources leave signatures in the isotopic abundances of CRs. Current models of Galactic CRs that consider supernovae (SNe) shocks as the main sites of particle acceleration cannot satisfactorily explain the higher ${\rm ^{22}Ne/^{20}Ne}$ ratio in CRs compared to the interstellar medium. Although stellar winds from massive stars have been invoked, their contribution relative to SNe ejecta has been taken as a free parameter. Here we present a theoretical calculation of the relative contributions of wind termination shocks (WTSs) and SNe shocks in superbubbles, based on the hydrodynamics of winds in clusters, the standard stellar mass function, and stellar evolution theory. We find that the contribution of WTSs towards the total CR production is at least $25\%$, which rises to $\gtrsim 50\%$ for young ($\lesssim 10$ Myr) clusters, and explains the observed $^{22}{\rm Ne}/^{20} {\rm Ne}$ ratio. We argue that since the progenitors of apparently isolated supernovae remnants (SNRs) are born in massive star clusters, both WTS and SNe shocks can be integrated into a combined scenario of CRs being accelerated in massive clusters. This scenario is consistent with the observed ratio of SNRs to $\gamma$-ray bright ($L_\gamma \gtrsim 10^{35}$ erg s$^{-1}$)  star clusters, as predicted by star cluster mass function. Moreover, WTSs can accelerate CRs to PeV energies, and solve other longstanding problems of the standard supernova paradigm of CR acceleration.
\end{abstract}


\section{Introduction}
CRs have been thought to be mostly accelerated by SNe through shocks running into in the ISM \citep{Grenier2015}. One of the main arguments for this is the energy requirement of Galactic CRs to maintain a steady CR luminosity. This scenario demands $2\pm1$ SNe explosions per century in our Galaxy, given that $\sim 10\%$ of SNe energy goes to CRs (\citealt{Diehl2006}). However, this standard scenario is known to bear several ailing problems (e.g., \citealt{Gabici2019}) and additional/complementary sources of CRs have been sought in the literature. One such problem concerns the abundance ratios of certain isotopes which are different from solar abundances and yet not secondary products. For example, the observed ratio of $^{22}{\rm Ne}$ to $^{20}{\rm Ne}$ in Galactic CRs (GCRs) is $5.3\pm 0.3$ times the solar value (\citealt{Wiedenbeck1981}; \citealt{Binns2008}), and it cannot be satisfactorily explained by SNe shocks in the ISM.

It has been suggested that this observed anomalous ratio in GCRs can be explained if SN shocks in superbubbles (SBs) played a major role in accelerating CRs because the gas inside a SB is rich in $^{22}{\rm Ne}$, ejected in the winds of massive stars (\citealt{Higdon2003}; \citealt{Binns2008}). This fits in with the mounting evidence of CRs being accelerated in star clusters, as predicted by \cite{Cesarsky1983}. A signature of CRs in the form of $\gamma$-radiation has been detected in massive star clusters such as Cyg OB2, which are too young ($\lesssim 3$ Myr) to have had SN \citep{Ackermann2011}. Moreover, $\gamma$-ray luminosities of these clusters is $\sim 0.1\%$ of the wind mechanical power ($L_w$) feeding the cluster \citep{Ackermann2011,Yang2018}, and therefore, it raises the possibility of a significant contribution from young star clusters in CR acceleration, even in the absence of SNe shocks.

However, shifting the arena of CR acceleration from SNe shocks in the ISM to star clusters have not quite yielded a better estimate of the Neon isotope ratio. For this, one needs to estimate the relative contribution of shocks due to stellar winds and SNe shocks towards CR acceleration. This ratio has so far been treated as a free parameter, empirically chosen to fit the observed Neon isotopic ratio (e.g., \citealt{Murphy2016}) in the absence of a rigorous theoretical calculation. In order to understand observed anomalous $^{22}{\rm Ne}/^{20}{\rm Ne}$ in CRs, several empirical models have been proposed (e.g., \citealt{Higdon2003}; \citealt{Binns2008}; \citealt{Prantzos2012}). However none of these models have discussed the shock energetics and corresponding source of upstream particles (stellar wind and SNe ejecta), which goes into the acceleration process. It has also been pointed out that the $^{22}{\rm Ne}$ yield used in previous studies to explain the observed ratio from SB scenario is likely an overestimate \citep{Prantzos2012}. As an alternative scenario, some studies proposed CR acceleration by SNe shocks in the wind of massive progenitor star. However, these models needs some fine tuning, such as the requirement that SN shocks should be effective only when the shock speed $\ge 1600$ km s$^{-1}$ (which excludes the possibilities of CR acceleration by SNe shock beyond free-wind region of the progenitor star, for detail see e.g., \citet{Prantzos2012}. Therefore, although the basic premise of solving the Neon isotope problem by invoking stellar winds in star clusters seems appealing, detailed estimates have not been available, and in addition, other avenues also appear bleak.

In this paper, we take a fresh look at the Neon isotope problem, using hydrodynamics of winds in star clusters, the latest stellar evolutionary yields (\citealt{Limongi2018}), with the standard stellar mass function. Recently using  numerical simulations, we have shown that termination shocks of stellar winds from massive stars can accelerate CRs to produce $\gamma$-ray luminosity of a similar magnitude, as well as explain the observed X-ray and synchrotron luminosities \citep{Gupta2018b}. In this paper, we investigate the implications of our model to explain Neon isotope ratio. We show that the observed $^{22}{\rm Ne}/^{20}{\rm Ne}$ ratio can be achieved if CRs are produced in massive star clusters by the {\it combined} effects of WTS and SN shocks. With 1-D numerical simulation we estimate the fraction of shock energy processed in wind and SNe ejecta for various acceleration scenarios. We find that for compact clusters, WTS can accelerate particles from the wind material before the onset of SNe. We also show that, in the case of  SN shock in wind of progenitor star (\citealt{Prantzos2012}), the reverse shock is as efficient as the forward shock, thereby accelerating both wind material and SN ejecta. We demonstrate that in this case both the forward and reverse shocks are energetically comparable. Therefore the acceleration of the ejecta material (rich in ${\rm ^{20}Ne}$) cannot be neglected as is usually done in the literature, and this poses a problem for the ${\rm ^{22}Ne/^{20}Ne}$ abundance ratio.

The implications of this calculation goes beyond the Neon isotope problem. We show that our results imply more than a quarter of the Galactic CRs being accelerated in WTSs of star clusters. The extent of the shocked wind region also allows us to draw important conclusions about the maximum energy of accelerated CRs. We further argue that SNRs and stellar winds in clusters are both linked to massive stars, and therefore the two sites of CR acceleration, namely, WTS and SNe shocks may be put together on a common platform, of CR acceleration in SBs. Isolated SNe remnants would merely represent the lower end of the star cluster mass function, where OB stars number less than two. These lines of argument paint an integrated scenario of CR acceleration that not only solves Neon isotope and other problems, but also rids the standard paradigm of acceleration in SNRs of its generic problems.

The paper is structured as follows. We first characterise and discuss the WTS in \S 2. The numerical set-up is described in \S 3. In \S 4, we present our main results on the relative contribution of WTS and SNe shocks and estimate the Neon isotope ratio. In \S 5, we draw attention to a few important implications of our calculations for Galactic CRs. Our findings are summarized in section \S 6.

\section{Wind termination shock (WTS)}
\subsection{Formation of the WTSs} \label{subsec:formation_wts}
\begin{figure*}
\centering
\includegraphics[height= 3.4in,width=6.5in]{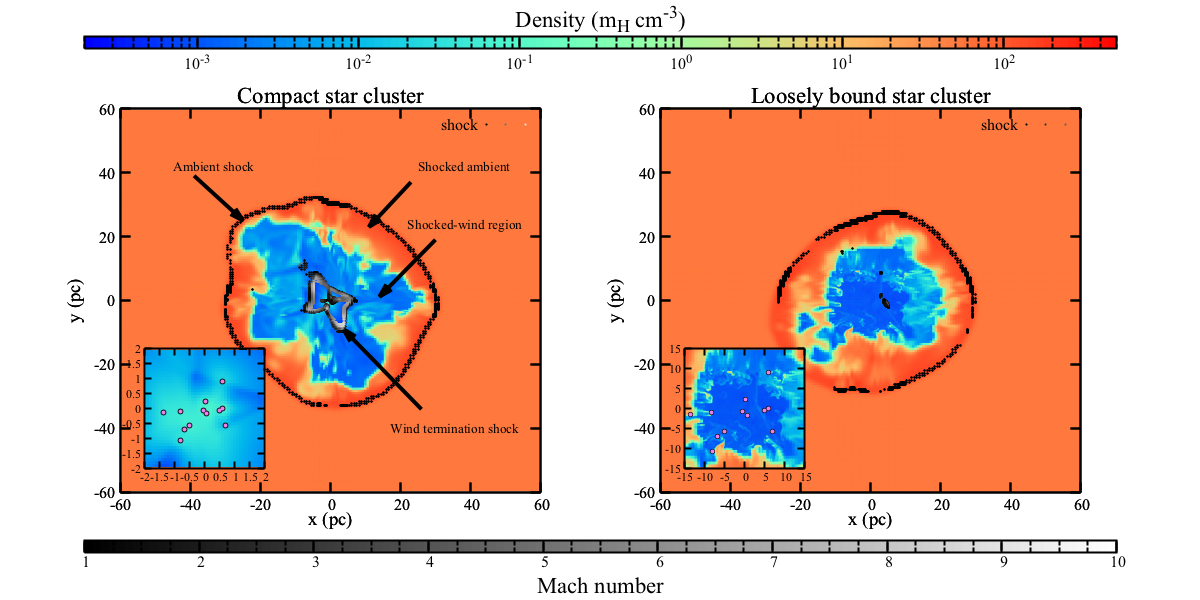}
\caption{Various diagnostics from $3$D star cluster simulations for a compact (left) and a loosely-bound cluster (right). Snapshots of density (top colour palette) and shock Mach numbers (bottom colour palette) in the $z=0$ plane at $2.9$ Myr are shown. Sub-plots show the zoomed-in view of the central few pc of the star cluster where magenta circles denote the locations of the stars after projecting them in the $z=0$ plane. Left and right panels show two different simulations that have identical set-up except for the core radius $R_{\rm c}$ of the cluster. For both left and right panels, the ambient shocks are weak and appear dark according to the bottom colour palette. The figure shows that a compact star cluster forms a WTS (left panel), and its Mach number is $\sim 5-10$, as shown by bright points.}
\label{fig:3dsim_den}
\end{figure*}
 Consider a star cluster in which the massive stars ($M_{\rm *} > 8$ $M_{\rm \odot}$) are mostly located within a radius $R_{\rm c}$. The total mechanical power launched by massive stars from this spherical region of radius $R_{\rm c}$ can be written as $L_{\rm w}= \dot{M} v^{2}_{\rm w}/2$, where $v_{\rm w}$ is the wind velocity and $\dot{M}$ is the total out-flowing wind mass per unit time. Interaction of the collective wind with the parent cloud forms an interstellar bubble. Near the core of this bubble, the wind expands adiabatically and its mass density is $\rho_{\rm w} = \dot{M}/(4\pi\,r^2\,v_{\rm w})$. A WTS forms at the location where the wind ram pressure ($P_{\rm ram}=\rho_{\rm w} v^{2}_{\rm w}$) balances the hot gas pressure ($P_{\rm in}$) in the bubble. Thus, the radius of the WTS w.r.t. the center of the star cluster is 
\begin{eqnarray} \label{eq:Rts}
R_{\rm ts} \approx \left(\frac{L_{\rm w}}{2\,\pi\,v_{\rm w}\,P_{\rm in}}\right)^{1/2}\ .
\label{eq:Rts}
\end{eqnarray}
This suggests that if a cluster is compact (i.e, $R_{\rm c} \ll R_{\rm ts}$) then a WTS can form. We have confirmed this by performing two $3$D simulations of star cluster of twelve massive stars (i.e., $N_{\rm OB} = 12$) with two different core radius $R_{\rm c}= 0.5$ pc and $5.0$ pc respectively. The results are shown in Fig. \ref{fig:3dsim_den} and for numerical set-up, see Appendix \ref{subsec:num3d}.

Left and right panels of Fig. \ref{fig:3dsim_den} show density snapshots in the $z=0$ plane at $2.9$ Myr for compact (left) and loosely bound (right) star clusters. The left panel shows that a compact cluster has formed a coherent WTS as shown by grey points (the shock Mach numbers are shown by grey/black points displayed in the horizontal colour palette). This can be understood as follows.

For this set-up, the total wind power of both clusters is $L_{\rm w} = N_{\rm OB} \times 10^{36}\,{\rm erg\,s^{-1}} \approx 1.2\times 10^{37}\, {\rm erg\,s^{-1}}$ and wind velocity is $v_{\rm w} = (2L_{\rm w}/\dot{M})^{1/2} \approx 2000\,{\rm km\, s^{-1}}$.
 Taking the hot gas pressure $P_{\rm in}\sim 10^{-10}\,{\rm dyne\,cm^{-2}}$ (as observed in SBs, see e.g., Table 7 in \citealt{Lopez2014}), from Eq. \ref{eq:Rts} we obtain $R_{\rm ts}=3.2$ pc. In the left panel, since the stars are distributed in a region smaller than $3.2$ pc (see the zoomed-in sub-plot), a WTS has formed\footnote{In this calculation, we have assumed that wind mechnical power of each star remains steady. In case of episodic winds, there will be internal shocks in the collective wind region but they will not affect the qualitative picture of the existence of WTS.}. In contrast, the right panel shows that the stars are distributed much beyond $3.2$ pc, and there is no coherent WTS. We have also confirmed this for the clusters of mass $\gtrsim 10^{3}\,M{\rm_{\odot}}$. Therefore, compact star clusters can form WTS.

\subsection{Mach number of WTSs}
In the case of compact clusters, the physical properties of the wind (e.g., density, velocity, and pressure profiles) in the region $r< R_{\rm ts}$ are similar to the model of \citet{Chevalier1985}. For $r\geq R_{\rm ts}$, there is a shocked wind region, which is separated from the ambient medium via a contact discontinuity (for details, see \citealt{Weaver1977}). In order to determine the Mach number of the WTS, we need to know the shock velocity and the upstream wind (free-wind) profile. In the lab frame, the WTS slowly moves outward (see e.g., Eq. 9 in \citealt{Gupta2018a}), the upstream velocity is the same as the wind velocity. The upstream sound speed can be obtained by using Table 1 in \citet{Chevalier1985}. This gives the Mach number of WTS as
\begin{eqnarray} \label{eq:Machana}
\mathcal{M} = \frac{v_{\rm w}}{c_{\rm s}} =  \frac{v_{\rm w}}{0.56 \dot{M}^{-1/2}\, L_{\rm w}^{1/2} \left(R_{\rm ts}/R_{\rm c}\right)^{-2/3}} \simeq 2.5\, \left(\frac{R_{\rm ts}}{R_{\rm c}}\right)^{2/3} .
\end{eqnarray}
This implies that a large separation between WTS and the cluster core leads to a large Mach number which increases with time as the termination shock moves out.

For a typical cluster of mass $10^{3}\,M{\rm_\odot}$
($N_{\rm OB}\approx 12$)
 Eq. (\ref{eq:Rts}) gives $R_{\rm ts}\approx 3.2\,{\rm pc}\, P^{-1/2}_{\rm in,-10}\,L_{\rm w, 37}^{1/2}$. The corresponding Mach number of WTS is 
\begin{equation} \label{eq:Mach}
\mathcal{M} \approx 6 \, \left(\frac{R_{\rm c}}{pc}\right)^{-2/3}\,\left(\frac{P_{\rm in}}{10^{-10}\, {\rm dyne\,cm^{-2}}}\right)^{-1/3}\,\left(\frac{L_{\rm w}}{10^{37}\, {\rm erg\,s^{-1}}}\right)^{1/3}\ .
\end{equation}
This suggests that compact clusters can have high Mach number WTS. In contrast, the outer shock Mach number is small as seen in Fig. \ref{fig:3dsim_den}.

Although these calculations refer to star clusters, Eq. (\ref{eq:Mach}) is also valid for the wind from a single star. In this case, $R_{\rm c}$ can be taken as the radius beyond which the wind becomes supersonic. This scenario is also applicable for bigger star clusters (i.e., core radius $>> R_{\rm ts}$) in which a global WTS may not form (as shown in the right panel of Fig. \ref{fig:3dsim_den}). In this case, individual stars can accelerate CRs in their WTS. We discuss these cases separately in section \ref{subsubsec:case3}.

\begin{figure*}
\centering
\includegraphics[height= 4.7in,width=5.3in]{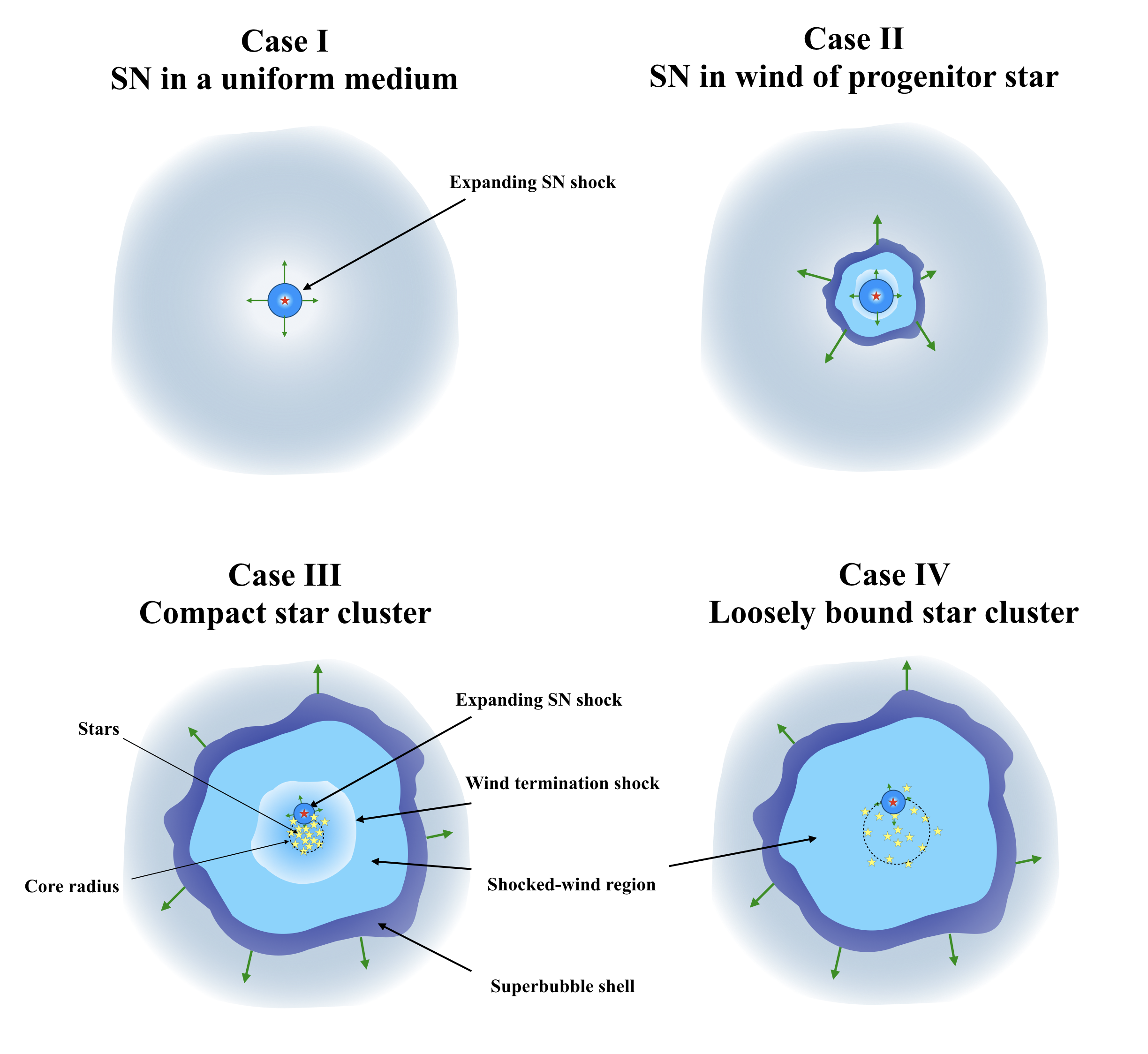}
\caption{Schematic diagram of four different CR acceleration scenarios.}
\label{fig:cr_cases_schematic}
\end{figure*}
\section{Numerical set-up} \label{sec:shockE}
The central result of the present work is to compare the energy efficiency in accelerating wind material, SNe ejecta, and ambient matter by the WTS and SNe shocks. For this purpose, we need to quantify the fraction of upstream energy that is encountered by wind material, SNe ejecta, and ambient matter. In order to estimate the energy processed at different shocks, we consider four different acceleration scenarios, which can be broadly classified into two categories: (1) isolated SN (sections \ref{subsubsec:case1} and \ref{subsubsec:case2}) and (2) star cluster (sections \ref{subsubsec:case3} and \ref{subsubsec:case4}). In the first case, SN shock expands either in the parent cloud or in the wind of the progenitor star. In the second case, the star cluster may be compact or loosely bound. We have labeled these four acceleration scenarios as Case I to Case IV (for a brief overview see Fig. \ref{fig:cr_cases_schematic}). For each of these cases, we discuss the energetics of various shocks with the help of $1$-D simulations.
\begin{figure*}
\centering
\includegraphics[height= 5.9in,width=5.6in]{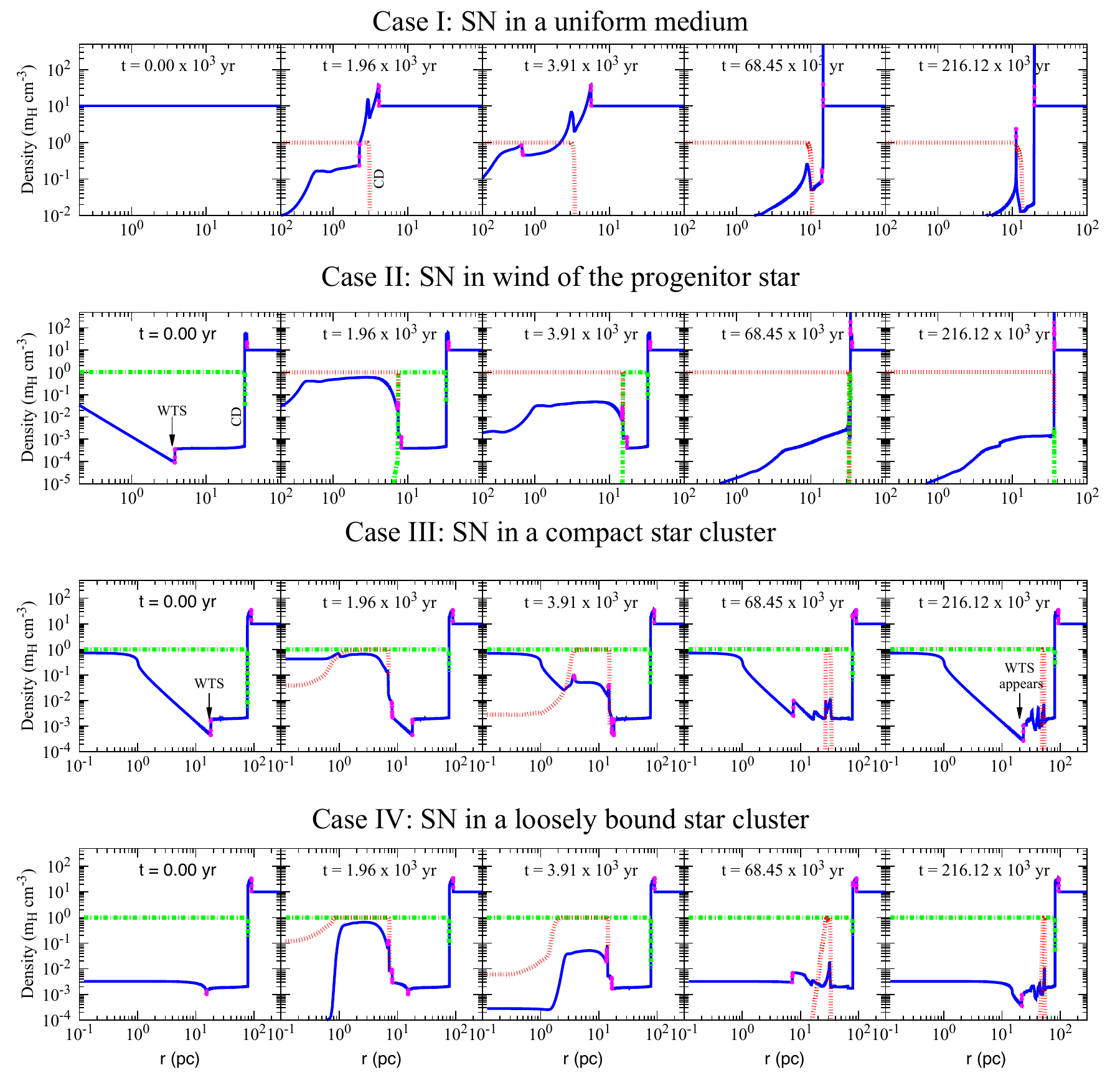}
\caption{Four different CR acceleration scenarios: in all panels, blue curve represents density profiles, where time $t=0$ denotes the epoch of SN explosion. Red and green curves display tracer of wind material and SN ejecta respectively. Four different acceleration scenarios, which are labeled by Case I to Case IV, are described as follows. Case I: explosion of an isolated star in a uniform medium ($t=0\Rightarrow t_{\rm dyn}=0$), Case II: explosion of an isolated star in wind of the progenitor star ($t=0 \Rightarrow t_{\rm dyn}=3.5$ Myr), Case III: SN explosion in a compact star cluster ($t=0 \Rightarrow t_{\rm dyn}=3.5$ Myr). Case IV: SN explosion in a medium made by a loosely bound star cluster ($t=0 \Rightarrow t_{\rm dyn}=3.5$ Myr).}
\label{fig:cr_cases}
\end{figure*}
\begin{figure*}
\centering
\includegraphics[height= 3.25in,width=6.4in]{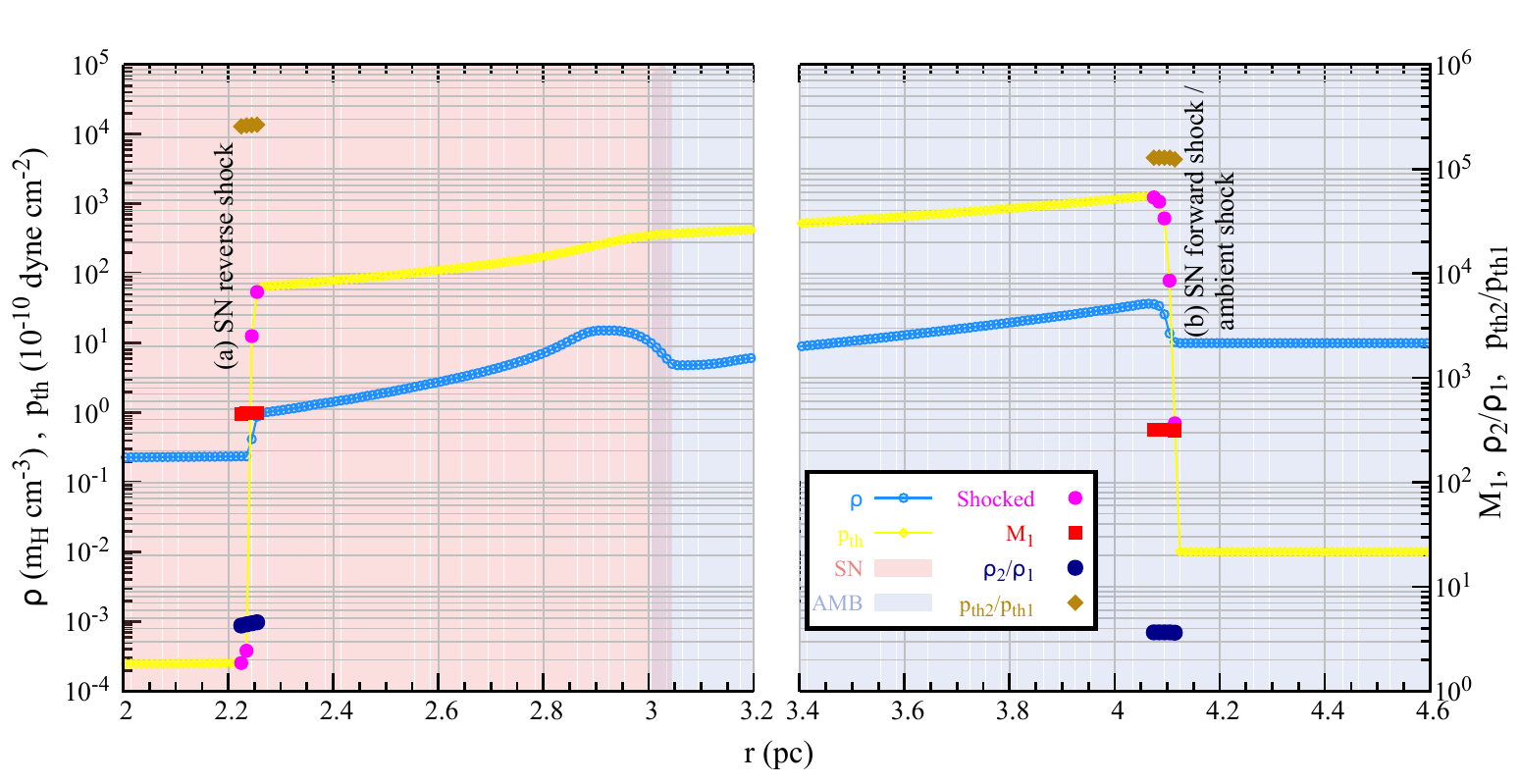}
\caption{Shock diagnostics for an isolated SN in a uniform medium (case - I). Left and right panels display the zoomed-in view of density (blue) and pressure (yellow) profiles of the blast wave near the reverse shock (left) and near the ambient forward shock (right) respectively at $t=1.96\times 10^{3}$ yr. The shock profiles shown here represent the second panel of case I in Figure \ref{fig:cr_cases}. The background colours of this figure display the tracer of SN ejecta (light-red) and ambient matter (light-steel-blue). The right axis shows the upstream Mach number $M_{\rm 1}$ (red squares), density jump $\rho_{2}/\rho_{\rm 1}$ (dark-blue circles) and pressure jump $p_{\rm 2}/p_{\rm 1}$ (brown diamond symbols). The right panel (for the forward shock) shows that $M_{\rm 1}\approx 300$, $p_{\rm 2}/p_{\rm 1}\approx 1.2\times 10^{5}$, and $\rho_{\rm 2}/\rho_{\rm 1}\approx 4$, which are consistent with our analytical estimates.}
\label{fig:c1shockpf}
\end{figure*}

We solve the standard Euler equations in $1$-D spherical geometry. We use uniformly distributed grids with spatial resolutions $\Delta r = 0.01$ pc for the Case I, II and $\Delta r = 0.04$ pc for the Case III, IV respectively.  In order to include the effect of stellar winds/SNe, we consider a spherical region of radius $r_{\rm inj}$ within which we inject mass and energy uniformly. Depending on the acceleration scenario that we wish to study, $r_{\rm inj}$ ranges from $0.05$ to $15$ pc. The SN shock is launched by injecting thermal energy $E_{\rm SN} =10^{51}\,{\rm erg}$ and mass $M_{\rm ej}=10\,M{_{\odot}}$ in a region of radius $0.05$ pc ($0.25$ pc) for cases I and II (cases III and IV). For cases II, III, and IV, the first SN occurs at $3.5$ Myr, which corresponds to the main sequence life time of $\sim 100\, M_{\rm \odot}$ star. The cases III and IV represent clusters of mass $10^{4}\, M_{\rm \odot}$ with two different core radii: $1$ pc and $15$ pc respectively. For these two cases, the time between two consecutive SN explosions is $\Delta \tau_{\rm SN} = \tau_{\rm cluster}/ N_{\rm OB}$, where $N_{\rm OB} \simeq 109$ and $\tau_{\rm cluster} \approx 30$ Myr is typical cluster lifetime. For all cases, we assume the initial ambient density $\rho=10\,m{\rm _H\,cm^{-3}}$, and pressure $10^{-12}\, {\rm dyne\,cm^{-2}}$. Radiative cooling has been included using a tabulated cooling function (\citealt{Sutherland1993}). We have used three passive scalars to distinguish wind, SN ejecta, and ambient matter. These passive scalars help us to identify the material(s) available in the upstream/downstream region of a shock. This is required to find the energy that goes into different types of material: wind, SN ejecta, and ambient matter.
\subsection{Shock energetics} 
Various steps in the analysis are described in sections below.
\begin{itemize}
\item Step 1 - Simulation output: Directly obtained from our hydro runs.
\item Step 2 - Shock detection: The following three conditions  are used to identify the shocked zones.
\begin{eqnarray} \label{eq:con1}
\nabla\cdot {\bf v}  & < &  0,\,\\
{\bf \nabla}p \cdot \Delta r/p  & > & \delta_{\rm threshold},\, {\rm and}\,\\
{\bf \nabla}T\cdot {\bf \nabla}\rho & > &  0\ .
\end{eqnarray}
Here $\rho, p,$ and $T$ are the density, pressure and temperature of the fluid.
The first condition selects compressed zones. The second condition sets a threshold in pressure jump ($\delta_{\rm threshold} = 0.5$), and the third conditions helps to avoid contact discontinuity. We have confirmed that these conditions robustly identify shocked zones.
\item Step 3 - Identifying upstream/downstream parameters: 
 For each shocked zone, the program compares density, pressure, and sound speed of $4$ to $8$ neighbouring zones on both sides of the shocked zone. 
Finally, it gives the density, pressure, and sound speed of the upstream and downstream regions.
\item Step 4 - Next we determine the following entities.
\begin{enumerate}
\item Density compression ratio: $\rho_{\rm 2}/\rho_{\rm 1}$.
\item Pressure jump: $p_{\rm 2}/p_{\rm 1}$
\item Upstream sound speed: $c_{\rm 1} = (\gamma\, p_{\rm 1}/\rho_{\rm 1})^{1/2}$, where  $\gamma = 5/3$.
\item Upstream Mach number: $M_{\rm 1} = \frac{1}{(2\gamma)^{1/2}}\left[\frac{p_{\rm 2}}{p_{\rm 1}}(\gamma+1) + (\gamma-1)\right]^{1/2}$
\item Mass flux: $\dot{\rho}_{\rm m} = \rho_{\rm 1}v_{\rm 1} = \rho_{\rm 1} (c_{\rm 1} M_{\rm 1}) = \rho_{\rm 2}v_{\rm 2} $.
\item Energy flux: $\dot{\rho}_{\rm e} = [(5/2)p_{\rm 1}/\rho_{\rm 1}+ v^{2}_{\rm 1}/2]\rho_{\rm 1}v_{\rm 1} =  [(5/2)p_{\rm 2}/\rho_{\rm 2}+ v^{2}_{\rm 2}/2]\rho_{\rm 2}v_{\rm 2}$
\end{enumerate}
\end{itemize} 
Finally, we obtain the total mass/energy flux that crosses the shock surface. This is estimated by multiplying the flux with the shock surface area. In numerical simulations, since the shock surface is made of more than one zone, we have calculated the shocked-zone averaged entities by using the following equations:
\begin{eqnarray} \label{eq:fluxA} 
\dot{m}_{\rm T}   = \frac{\displaystyle \sum_{i} \dot{\rho}_{\rm m, i}\, A_{\rm i}}{\displaystyle \sum_{i} Z_{\rm i}}\, ,\ \,{\rm and}\, \ \,  
\dot{e}_{\rm T}   = \frac{\displaystyle \sum_{i} \dot{\rho}_{\rm e, i}\, A_{\rm i}}{\displaystyle \sum_{i} Z_{\rm i}}
\end{eqnarray}
where $i$ denotes the effective shocked zones where the analysis is performed, $A_{\rm i} = 4\pi r^{2}_{\rm i}$ is the shock area, and $\sum_{i} Z_{\rm i}$ is the total number of zones in a shock surface (typically $\sum_{i} Z_{\rm i}\approx 4-8$). The net flow of mass/energy through the shock surface are calculated using
\begin{eqnarray}\label{eq:fluxAdt}
m_{\rm T}   = \int_{t}dt\, \dot{m}_{\rm T} \, ,\ \,{\rm and}\, \ \, 
e_{\rm T}    = \int_{t}dt\, \dot{e}_{\rm T} \ .
\end{eqnarray}
Note that, the computational domain may contain more than one shock surface encountered by the same material at different locations. In this case, we have to find zone averaged entities for each shock surface separately (e.g., $\sum_{\rm shock} \dot{m}_{\rm T}$). In order to do this, we define a critical distance $r_{\rm cri} = 8 \Delta r$ (i.e., the length of shock analysis domain), to specify the minimum distance between two shock surfaces. When the separation between two neighbouring shocked zones is smaller than $r_{\rm cri}$, they are considered as the part of the same shock surface.
\section{Results} \label{subsec:results}

The results of our shock analysis program for the cases I, II, III, and IV are discussed below. For a brief overview of different cases, see Fig. \ref{fig:cr_cases_schematic}. 

Fig. \ref{fig:cr_cases} shows the density profiles at five different epoch for four different cases. Panel I shows the classical results of SN in a uniform medium (Case I). The magenta dots show the position of forward and reverse shocks of the blast wave. The red curves show that the SN ejecta reaches up to the contact discontinuity (hereafter, CD). For Case II, the first panel shows the density profile just before the SN, where position of the WTS is marked by an arrow. The green curves show that wind material reaches up to the CD. When the SN shock reaches WTS and collides with it, the WTS disappears. In contrast, in Case III, the WTS appears again due to the winds from remaining stars. For Case IV, although we see shocked zones at WTS, it is weak compared to that of case III because the separation between WTS and core radius is small (see e.g., Eq. \ref{eq:Machana}). Note that, in the rightmost panel of Case I, II ($t=216.12\times 10^{3}$ yr), the interior of the bubble contains SN ejecta. In contrast, in the Case III and Case IV, SN ejecta is accumulated near the swept up ISM (shell) and the interior of the bubble is filled with wind material. 

\subsection{Shock energetics}
We have described each of these cases in detail in the following sections. Note that, each section contains two figures. The first figure shows zoomed-in shock profiles (i.e., Figs. \ref{fig:c1shockpf}, \ref{fig:c2shockpf}, \ref{fig:c3shockpf}, and \ref{fig:c4shockpf}), where the blue 
 curves stand for density and yellow curves for pressure. The second figure in each section shows the time evolution of mass/energy of the material that is swept up by the shock (i.e., Figs \ref{fig:c1}, \ref{fig:c2}, \ref{fig:c3}, and \ref{fig:c4}), where green/cyan stands for wind material, red/light-red denotes SN ejecta and dark-blue/light-steel-blue stands for ambient matter.

\subsubsection{Case - I: SN in a uniform medium}  \label{subsubsec:case1}
For a SN in a uniform medium, we have used the blast wave solution (e.g.,  \citealt{Truelove1999}) to check our analysis program. For our set-up (i.e., $E_{\rm SN}=10^{51}\, {\rm erg}$, $M_{\rm ej}=10\, M_{\rm \odot}$, and $\rho = 10\,m{\rm_H\,cm^{-3}}$), at $t=1.96\times 10^{3}\, {\rm yr}$, $R\approx 4.2$ pc and $\dot{R}\approx 840\,{\rm km\,s^{-1}}$. Therefore, we expect the shock Mach number $M_{\rm 1}\approx 840/3= 280$, pressure jump $p_{\rm 2}/p_{\rm 1} = (5/4)\,M_{\rm 1}^2 \approx 10^{5}$, and compression ratio $\rho_{\rm 2}/\rho_{\rm 1} \approx 4$. Our shock analysis program confirms this, as shown in Fig. \ref{fig:c1shockpf}.

\begin{figure*}
\centering
\includegraphics[height= 3.6in,width=6.6in]{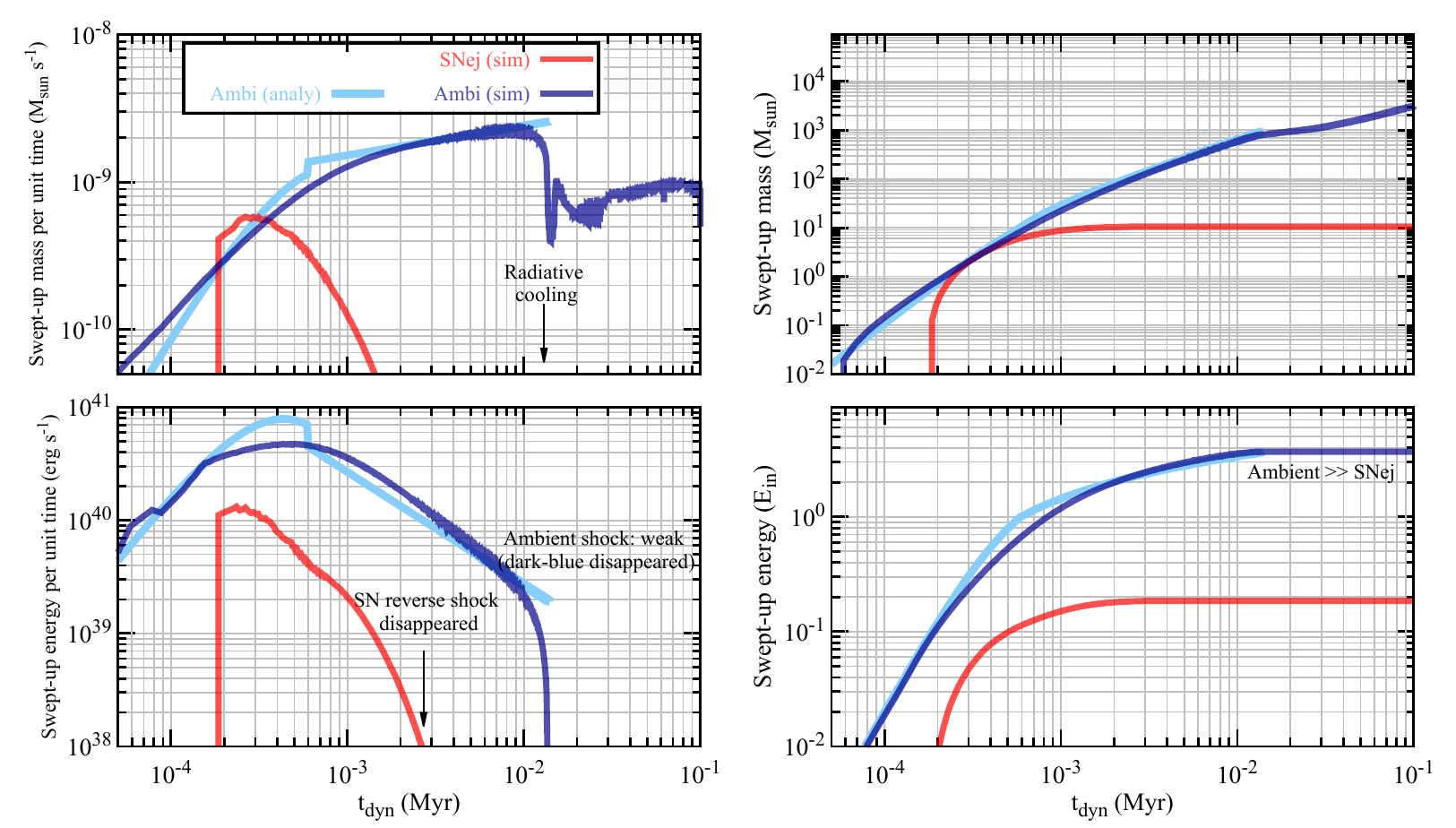}
\caption{Various diagnostics from an isolated SN simulation (Case I). Left panels display the net mass/energy that passes through the shock surface per unit time. Right panels display the time integrated entities corresponding to the left panels. In the bottom right panel, the vertical axis is normalized w.r.t. the input energy (i.e., $E_{\rm in}=10^{51}\,{\rm erg}$). Sky-blue solid curves display analytic predictions (representing forward shock), and blue (ambient)/red (SN ejecta) curves show the results from our analysis. Top right panel shows that the mass swept-up by the reverse shock is equal to ejecta mass $M_{\rm ej}\approx 10\, M{\rm _\odot}$ as expected. Right bottom panel shows that the reverse shock is energetically sub-dominant compared to the forward (ambient) shock of the blast wave. In this case, acceleration of the ambient matter is energetically dominant, with only 5\% of the total shock energy processed by the SN ejecta. Note that, the energy swept up in the ambient medium is larger than unity because our calculation does not include the loss of upstream energy due to deceleration of the shell (see e.g., Eq. $5$ in \citealt{Dermer2013}).}
\label{fig:c1}
\end{figure*}


Fig. \ref{fig:c1shockpf} shows the zoomed-in density (sky-blue) and pressure (yellow) profiles of a blast wave near the reverse shock (left) and the forward shock (right). The magenta points show the shocked zones. Two background colours, i.e., light-red and light-steel-blue, represent the tracers of SN ejecta and ambient matter respectively. In the left panel, the transition between these two colours represent the location of the contact discontinuity. 


 Top/bottom panels of Fig. \ref{fig:c1} show various diagnostics of mass/energy that passes through the shock surfaces calculated using Eqs \ref{eq:fluxA} and \ref{eq:fluxAdt}. Blue and red curves stand for ambient matter and SN ejecta respectively. Sky-blue solid curves represent the analytical predictions. 
 The bottom right panel shows that the reverse shock energy is smaller than the forward shock by a factor of ten. Therefore, the acceleration of SN ejecta is energetically not preferred in this scenario. In this case, the shock energy is mainly encountered by the ambient matter.

\subsubsection{Case - II:  SN in wind of the progenitor star}  \label{subsubsec:case2}
\begin{figure*}
\centering
\includegraphics[height= 3.4in,width=6.4in]{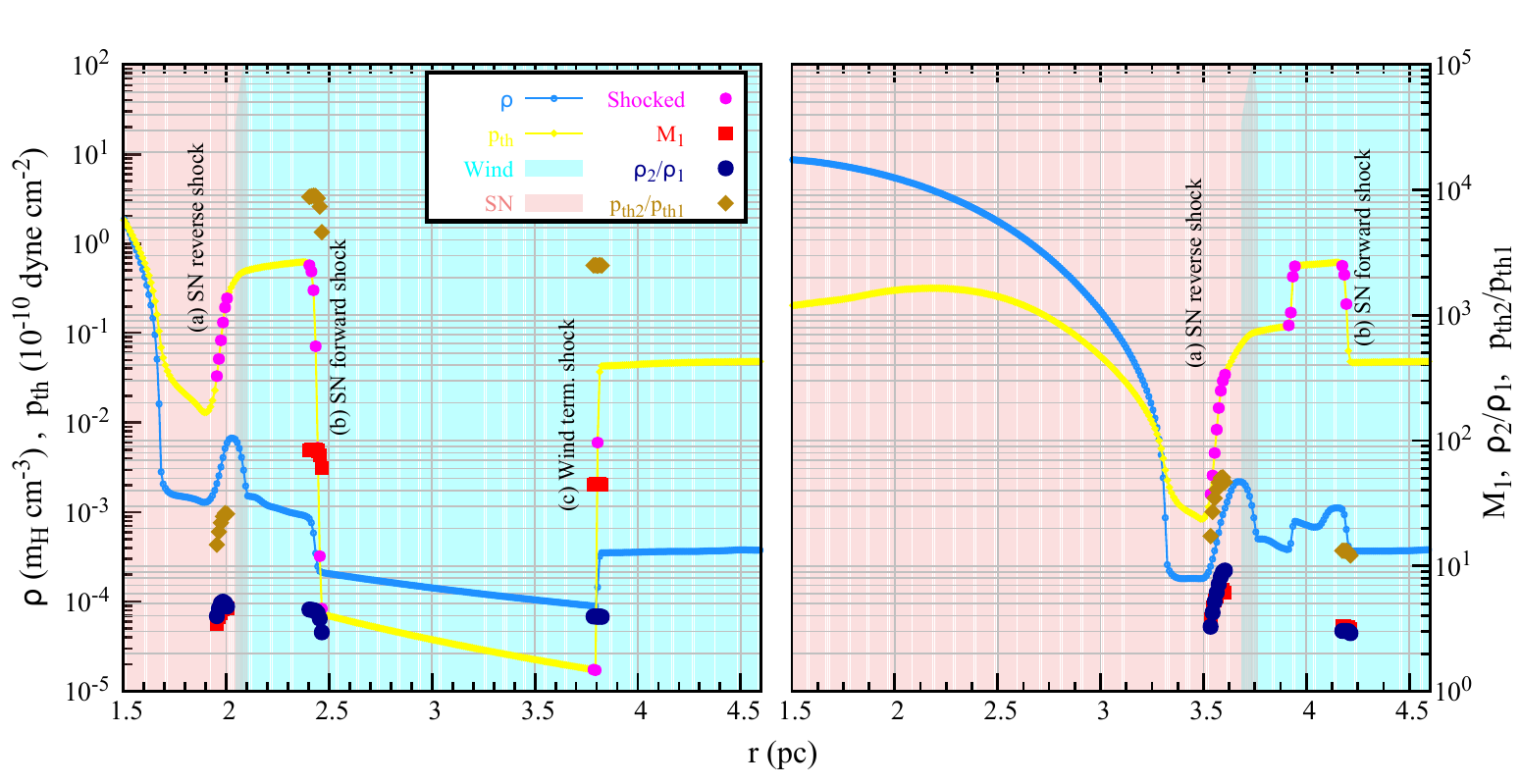}
\vspace{-1em}
\caption{Left and right panels display the zoomed-in view of density (blue) and pressure (yellow) profiles near the WTS at two different times separated by a time interval of $\approx 300$ yr. The left panel shows a phase when SN shock is propagating in free wind region and the right panel shows just after SN shock hits the WTS. The right axes represent upstream Mach number $M_{\rm 1}$ (red squares), density jump $\rho_{2}/\rho_{\rm 1}$ (dark-blue circles), and pressure jump $p_{\rm 2}/p_{\rm 1}$ (brown diamond symbols) obtained from the analysis program. The tracer of wind material and SN ejecta are shown by background colours (light-cyan for wind and light-red for SN ejecta as in Fig. \ref{fig:c1shockpf}). Left panel contains three distinct shock surfaces: (starting from left) (a) SN reverse shock (sweeping up SN ejecta), (b) SN forward shock (wind material), (c) wind termination shock (wind material). In this case the wind material encounters {\it two different shock surfaces} (i.e., shock - b, c) whereas the SN ejecta encounters a single shock surface (shock - a). As soon as the blast wave forward shock collides with the WTS (i.e., collision between shock b with shock c), the WTS disappears (see the right panel, shock - c is absent). Comparison of Mach numbers (red squares) between left and right panels near the shock (a) shows that the SN reverse shock becomes stronger after the SN forward shock and WTS collide. 
}
\label{fig:c2shockpf}
\centering
\includegraphics[height= 3.4in,width=6.4in]{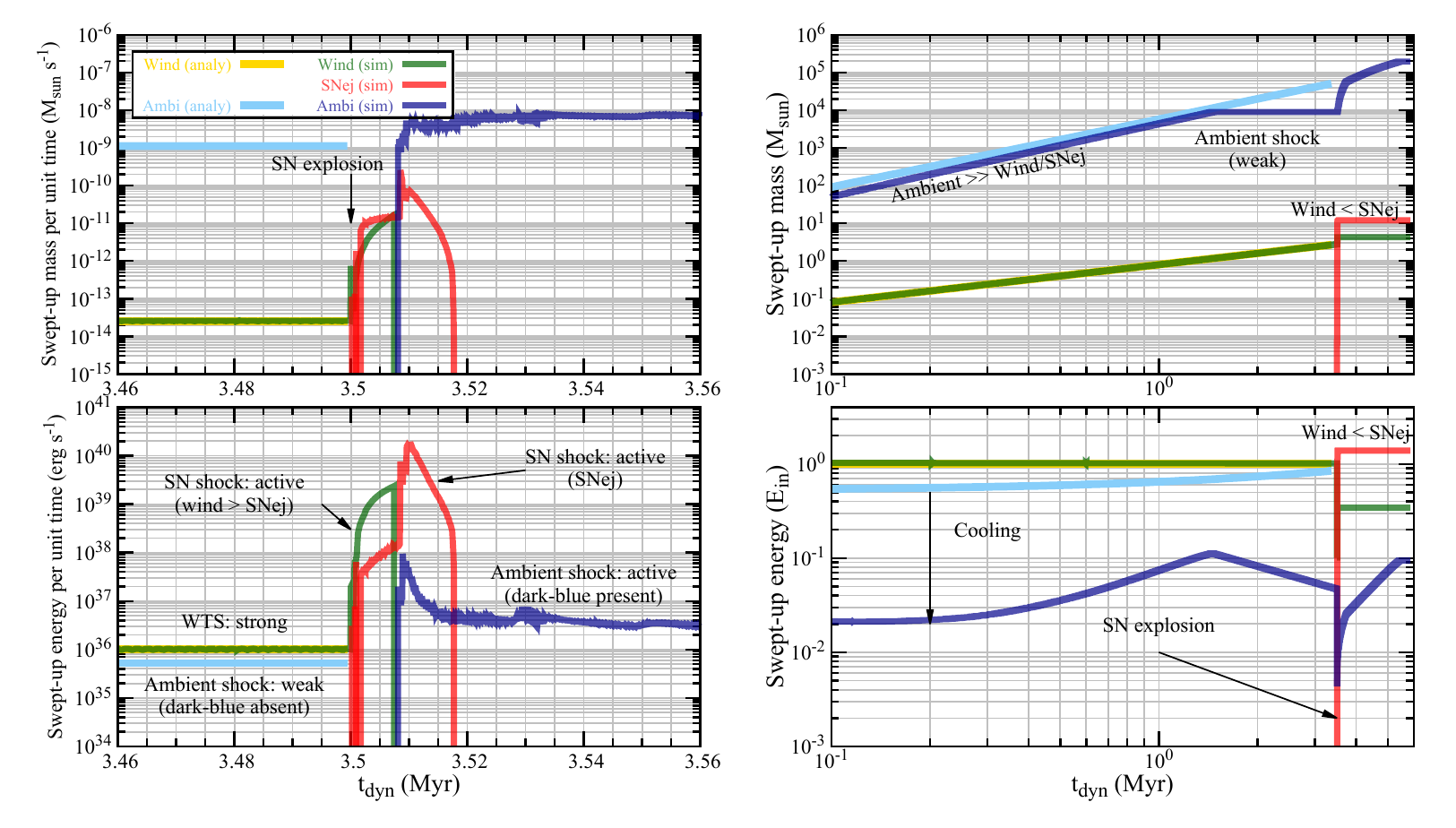}
\caption{Shock diagnostics from an isolated SN in wind of progenitor star (top/bottom panel: mass/energy). Unlike Case-I, in this model, we have included stellar wind until $3.5$ Myr when the star has exploded. Left panels shows the zoomed-in view of $3.46-3.56$ Myr, during the epoch of SN explosion. They display the net mass/energy that passes through the shocked surface per unit time (i.e., $\dot{m}_{\rm T}\,\dot{e}_{\rm T}$ using Eq. \ref{eq:fluxA}). Right panels display the time-integrated mass (top) and energy (bottom) i.e., $m_{\rm T} $ and $e_{\rm T}$ (using Eq. \ref{eq:fluxAdt}). In the bottom right panel, the vertical axis is normalized w.r.t. the total input energy until that epoch. In all panels, green, red, and blue curves denote wind, SN ejecta, and ambient matter respectively. Yellow/sky-blue solid curves display the expected results from analytic calculation (shown only for WTS and ambient shock).}
\label{fig:c2}
\end{figure*}

In this case, the stellar wind of the progenitor star has been included. Before the SN explosion, interaction of the wind with the ambient medium forms a wind bubble. This wind bubble has four distinct regions: (i) free wind, (ii) wind termination shock, (iii) shocked wind region, and (iv) swept up ambient shell (\citealt{Weaver1977}). In this case, the WTS can accelerate the wind material even before the SN. When SN occurs, the blast wave moves through these four regions mentioned above. We show the shock profiles in Fig. \ref{fig:c2shockpf}.

In the WTS rest frame the upstream fluid moves with velocity $v_{\rm w}$, and therefore, the mass and energy flux that pass through the WTS are roughly equal to the stellar wind mass-loss rate (i.e., $\dot{m}_{\rm T} = \dot{M}=8\times 10^{-7}\, M{\rm _\odot\,yr^{-1}}$) and wind power ($\dot{e}_{\rm T} =  L_{\rm w} = 10^{36}\, {\rm erg\,s^{-1}}$), as shown in the left of Fig \ref{fig:c2}. The shock evolution at different epochs are described below.

\begin{itemize}
\item The WTS sweeps up stellar wind material until $3.5$ Myr (green curves in the left panel). It also confirms the analytic prediction (yellow curve). 

\item  After SN explosion, the forward shock of the blast wave moves in the free wind region. In this phase, the wind material (green) is swept up by both WTS and SN forward shock. This can be seen from the sudden rise in the green curve at $3.5$ Myr. However, this phase lasts for a short time, typically $\approx 900$ yr from the epoch of explosion. The Galatic CR acceleration paradigm that considers SN blast wave in the wind of massive stars mostly focuses on this phase (e.g., \citealt{Prantzos2012}; \citealt{Biermann1993}). However, there are phases described below, which are also important and should not be neglected.

\item When the SN forward shock reaches the WTS, it collides with WTS and the WTS disappears. At this moment, a reflected shock and a transmitted shock are formed. The (transmitted) SN forward shock moves through the hot shocked-wind region where it sweeps up the wind material. During this time, the shock energy processed by the wind material is larger than that by the SN ejecta (in the bottom left panel, the green curve is above the red curve). This phase continues as long as the SN forward shock does not reach the CD of wind bubble (until $\approx 7500$ yr from the epoch of the SN explosion).

\item When (transmitted) SN forward shock reaches the CD, the wind material accumulates near the CD. Acceleration of wind material stops when the SN shock collides with CD. This occurs after $\approx 7500$ yr from the epoch of SN explosion (see green curves in left panels disappear after $\approx 3.5\,{\rm Myr} +7500$ yr).

\item The collision between the SN forward shock with shocked ambient medium again forms a transmitted shock and a reflected shock. The transmitted shock moves through the ambient medium. This can be noticed in the left panels where dark-blue curves suddenly appear after $3.5\,{\rm Myr} +7500$ yr.

\item The reflected shock and SN reverse shock moves towards the center of the explosion and continues to sweep up the SN ejecta (see the sudden rise in red curves at $\approx 3.5075$ Myr). They reach the point of explosion at $\approx 3.5\times 10^{6} + 2\times 10^{4}$ yr and both shocks disappear. 
\end{itemize}

Therefore, in this scenario, the strong shocks remain in bubble for $\approx 3.5\times 10^{6} +2\times 10^{4}$ yr, much longer than in an isolated SN. The top right panel shows that the swept up ejecta mass $\approx 10\, M_{\rm ej}$ (red), as expected. Comparison of red and green curves in the bottom right panel shows that the shock energy encountered by SN ejecta is {\it larger} than the wind/ambient material, a point usually missed in discussions of CR acceleration by SN shocks in massive stellar wind.

\begin{figure*}
\centering
\includegraphics[height= 3.4in,width=6.6in]{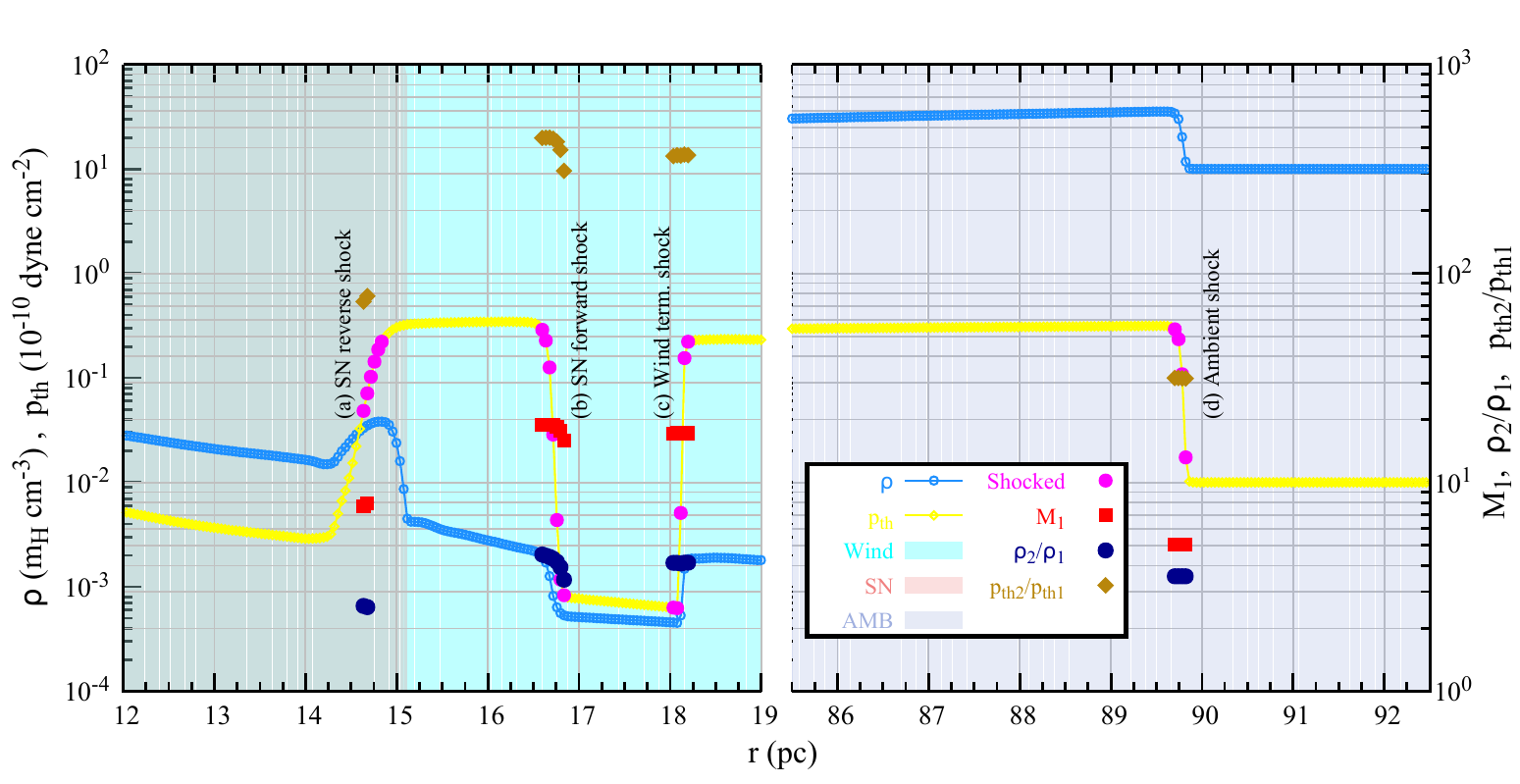}
\caption{The zoomed-in view of density (sky-blue) and pressure (light yellow) profiles of a SB near the WTS (left) and ambient shock (right) respectively at $t\approx 3.504$ Myr (i.e., immediately after the first SN). The used colour codes are identical with Figs \ref{fig:c1shockpf} and \ref{fig:c2shockpf}. The figure contains four shocked regions: (starting from left) (a) SN reverse shock ($M_{\rm 1}\approx 7$; sweeps up wind + SN), (b) SN forward shock ($M_{\rm 1}\approx 20$; sweeps up wind), (c) WTS ($M_{\rm 1}\approx 18$; sweeps up wind), and (d) ambient shock ($M_{\rm 1}\approx 5$; sweeps up ambient material). The figure shows that wind material is swept up at three locations: (a), (b), and (c), whereas the SN ejecta is swept up only at (a). Therefore, in the global energy budget, one can expect the acceleration of SN ejecta to be energetically sub-dominant compared to the wind material.}
\label{fig:c3shockpf}
\centering
\includegraphics[height= 3.6in,width=6.6in]{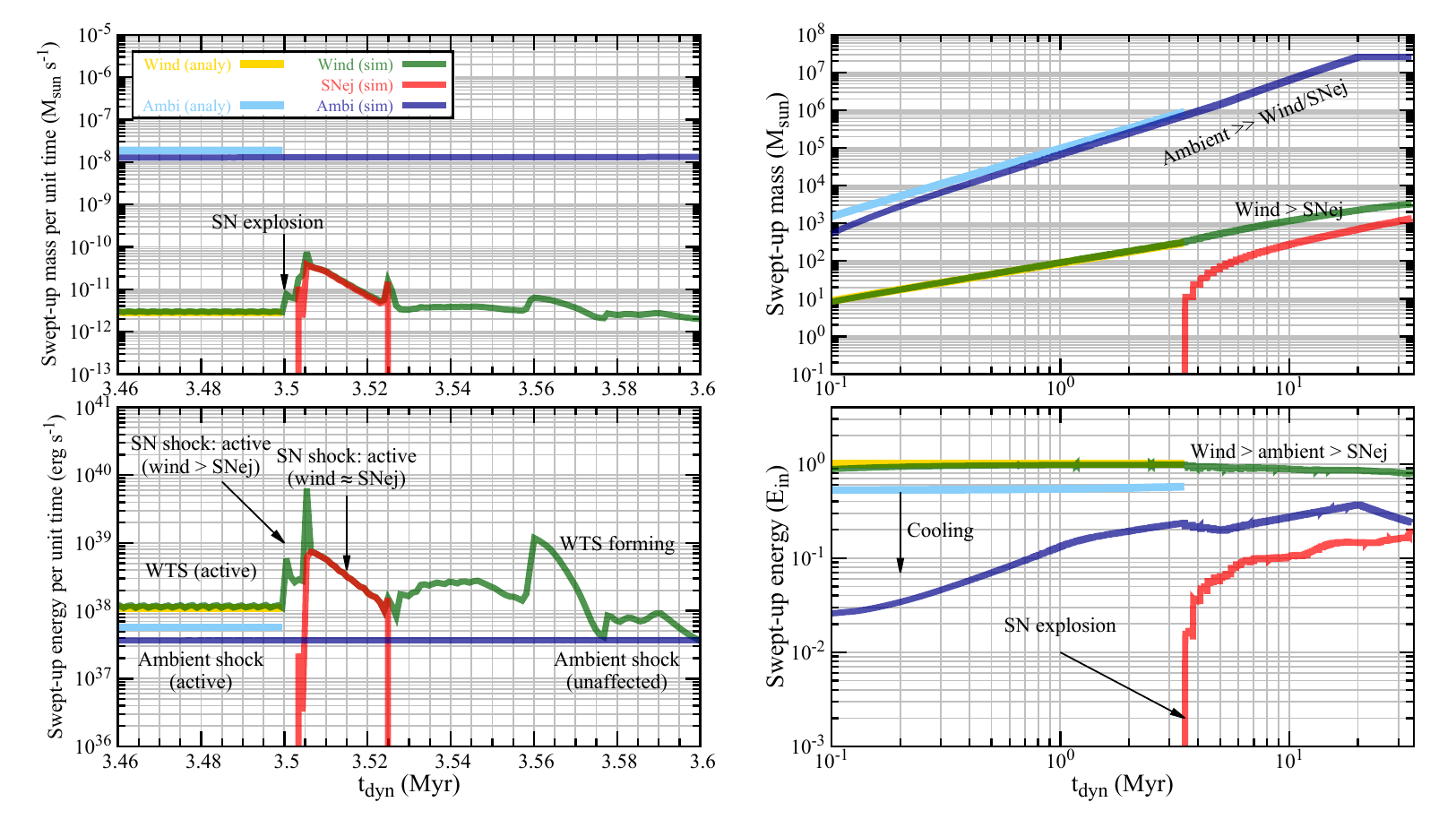}
\caption{Shock diagnostics of a compact star cluster (top panel: mass and bottom panel: energy), similar to  Figs \ref{fig:c1} and \ref{fig:c2}.
In the bottom right panel, comparison of red and green curves (dashed/solid) shows that the net shock energy going into SN ejecta is {\it smaller} than that of the wind material by a factor of $\sim 10$ at $\gtrsim 10$ Myr.}
\label{fig:c3}
\end{figure*}

\subsubsection{Case - III: Compact star clusters} \label{subsubsec:case4}
In this case, the stars are close to each other and the wind bubbles of individual stars overlap and generate a collective wind which forms a WTS. 
This WTS can accelerate the wind material, similar to Case II. When the first SN occurs in the cluster, the forward shock of the blast wave moves outward by sweeping up the wind material. In this case, the free wind is denser than that of a single star (i.e., Case II), and therefore, the forward shock of the blast wave spends a longer time in the free wind region.

The shock profiles at $3.5$ Myr $+$ $4\times 10^{3}$ yr are shown in Fig. \ref{fig:c3shockpf}. Left and right panels represent the zoomed-in shock profiles near the WTS and ambient shock respectively. The profiles look similar to Fig \ref{fig:c2shockpf}; however, in this case the SN reverse shock (labeled shock - a) also sweeps up the wind of remaining stars. This can be noticed from the background colour in the left panel of Fig. \ref{fig:c3shockpf} where the mixing between light-red (SN ejecta) and cyan (wind material) has produced a mixed colour (light-grey).

Fig. \ref{fig:c3} shows the shock energetics of wind material, SN ejecta, and ambient matter as described below.

\begin{itemize}
\item When the SN forward shock moves through the free wind region, it sweeps up wind material. See the left panels of Fig. \ref{fig:c3} where green curves suddenly rise at $3.5$ Myr. In this phase, the SN reverse shock is not strong (in the bottom left panel, the red curve at $\approx 3.5$ Myr is below the green curve).

\item The peak in the green curves at $3.505$ Myr represents the epoch of collision between the SN forward shock and WTS.
This collision produces a transmitted shock and a reflected shock, similar to Case II. However, unlike Case II, the transmitted shock that moves in the shocked-wind region may not reach the CD of the SB. This is because of the fact that the size of the bubble is bigger than Case II, the SN forward shock becomes subsonic. Therefore, the shocked ISM shell can remain unaffected. This is shown by dark-blue curves in the left panels of Fig \ref{fig:c3}.

\item The SN reverse shock keeps moving toward the center of the bubble. However, due to the winds of remaining stars, it cannot reach the central region. Instead, the winds of other stars push the SN reverse shock outwards. In this phase, the shock energy is equally shared by the wind material and SN ejecta (red curves fall on the top of green curves). This phase ends when the SN ejecta accumulate near the CD of the SB (red curves disappear at $\approx 3.5{\,\rm Myr}+ 2.5\times 10^{4}$ yr). By this time, the WTS forms again and the acceleration of wind material continues.

\item This chain of events repeats as long as the energy deposited by the wind is larger than that of SNe. When the number of SN events in the cluster becomes large, the WTS becomes weak. After that, SNe shocks accelerate both wind material and SNe ejecta. 
\end{itemize} 

Right bottom panel of Fig. \ref{fig:c3} shows that the fraction of the input energy encountered by the wind material is larger than the SN ejecta by a factor of $\gtrsim 6$. Therefore, in this scenario, the acceleration of {\it wind material is energetically dominant}.
  
\subsubsection{Case - IV: Loosely bound star clusters} \label{subsubsec:case3}
For loosely bound star clusters, a global WTS may not form. However, the individual stars can have their own WTSs. For a detailed investigation, we need to focus on two different length scales: (i) a global length scale (SB; $\gtrsim 10$ pc),  and (ii) a local length scale that focuses on the individual stars ($\lesssim 1$ pc).



\begin{figure*}
\centering
\includegraphics[height= 3.4in,width=6.5in]{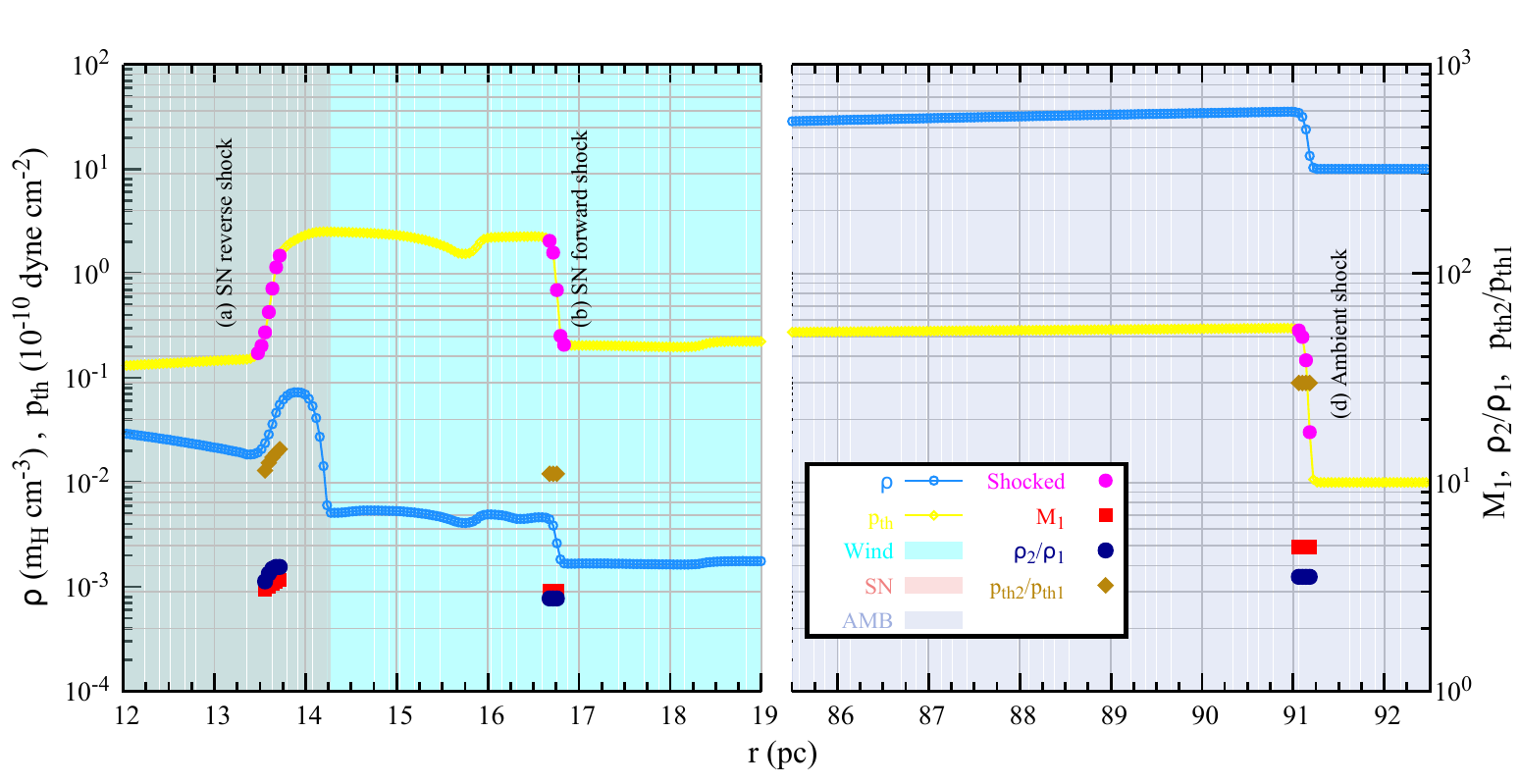}
\caption{The zoomed-in view of density (sky-blue) and pressure (light yellow) profiles of a SB near the SN forward/reverse shocks (left) and ambient shock (right) respectively at $t\approx 3.504$ Myr (the same epoch as in Fig \ref{fig:c3shockpf}). Unlike Case III (Fig. \ref{fig:c3shockpf}), this figure contains only three shock surfaces: (starting from left) (a) SN reverse shock ($M_{\rm 1}\approx 4$; sweeping up wind+SN), (b) SN forward shock ($M_{\rm 1}\approx 3$; sweeping up the wind material), and (d) ambient shock ($M_{\rm 1}\approx 5$; sweeping up ambient matter), i.e., a global WTS  (shock - c)  is missing. The figure shows that wind material is swept up at the shocks (a) and (b), whereas the SN ejecta is swept up only at (a). Therefore, the shock energy encountered by wind material can be larger than the SN ejecta.}
\label{fig:c4shockpf}
\centering
\centering
\includegraphics[height= 3.6in,width=6.6in]{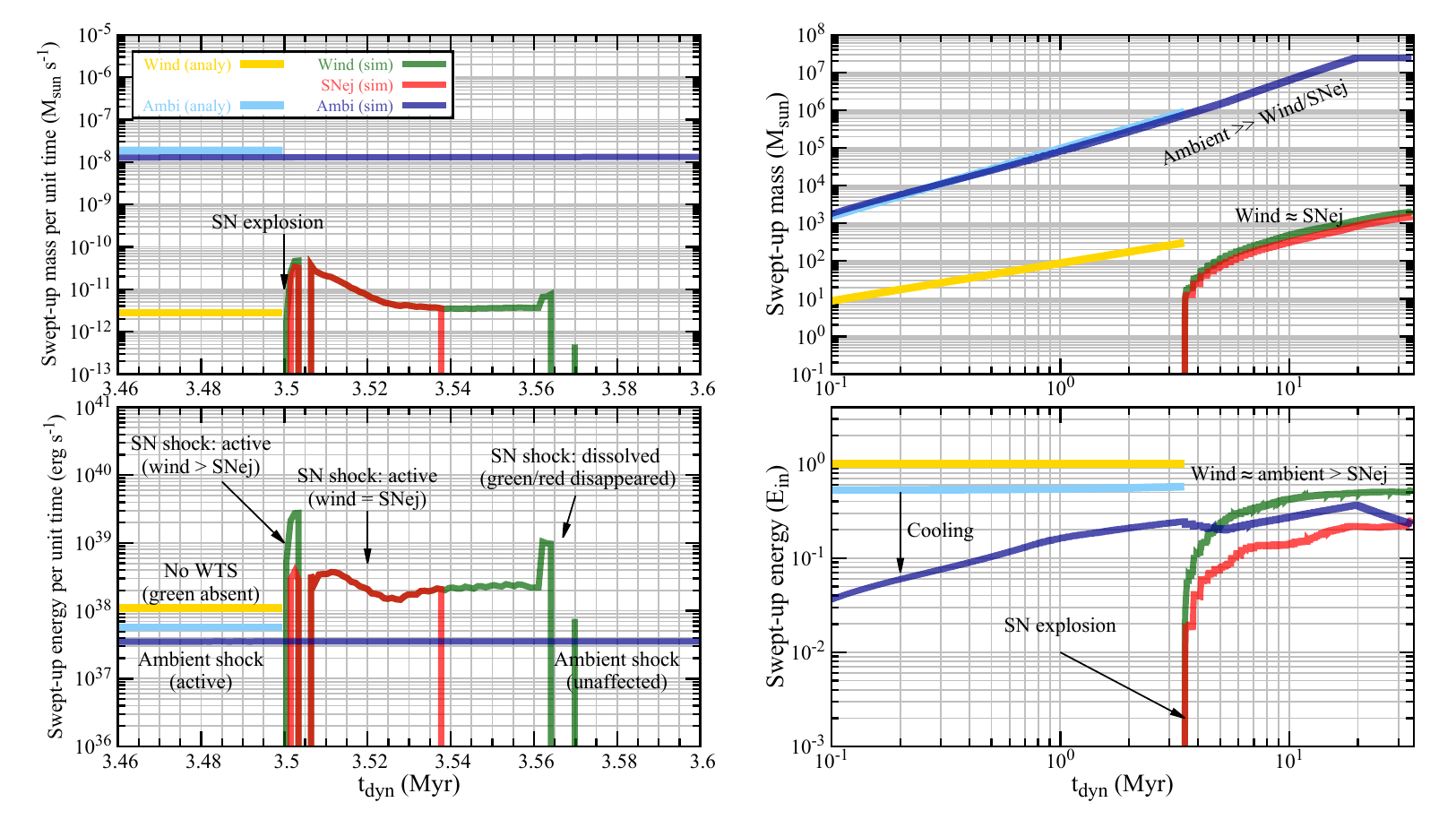}
\caption{Shock diagnostics of a loosely bound star cluster. In this case, the strong shocks appear after the first SN. The description of this figure is similar to Fig. \ref{fig:c4}. Comparison of red and green curves in the right bottom panel shows that the shock energy going into SN ejecta is smaller than that into wind material by a factor of $\sim 2$.}
\label{fig:c4}
\end{figure*}

In the global length scale, strong shocks appear only when SNe occur in the cluster. The zoomed-in density profiles near the SN forward/reverse shock and ambient shock are shown in Fig. \ref{fig:c4shockpf}. From the epoch of a given SN explosion, since the blast wave moves in a hot bubble, the shock Mach number is small compared to the case of compact star cluster (Case III). This can be seen by comparing the red squares of Figs. \ref{fig:c3shockpf} and \ref{fig:c4shockpf}.

The shock energetics are shown in Fig. \ref{fig:c4} and described below:
\begin{itemize}
\item In the left panels of Fig. \ref{fig:c4}, green curves are absent until $3.5$ Myr because WTS is absent. The wind material gets accelerated (green curves appear after $3.5$ Myr) after the first SN.

\item The wind material is swept up by both the forward and reverse shocks of the SN blast wave, similar to Case III. In this phase, the shock energy mostly goes into wind material (the green curves are above the red curves until $3.5{\rm Myr} + 5000$ yr).

\item Since the bubble is made of hot plasma, the SN forward shock disappears before it can reach the CD of the SB. The SN reverse shock remains active and moves toward the central region of the cluster. The wind of remaining stars push the SN ejecta and accumulate them near the CD of the SB. In this phase, the SN reverse shock energy equally goes to the wind material and SN ejecta (the red curves fall on the top of green curves until ($3.5$ Myr $+4\times 10^{4}$ yr), similar to Case III. 

\item When the SN reverse shock completely sweeps up the SN ejecta, the red curves disappear ($3.5$ Myr $+4\times 10^{4}$ yr). However, the SN reverse shock continues to sweep up the wind material till $3.5$ Myr $+6\times 10^{4}$yr and then it disappears (green curves disappear).

\item The above scenario repeats itself whenever a star explodes in the cluster.
\end{itemize}

The bottom right panel of Fig. \ref{fig:c4} shows that the shock energy going into SN ejecta is smaller than that into wind material by a factor of $\sim 2$ (unlike the factor of $\gtrsim 6$ for compact clusters).

There are some additional physical processes at local length scale of a loosely bound star cluster. In this case, a star is located in a hot medium made by the winds of other stars. Here, the ambient gas (inside the hot bubble) is mostly dominated by the wind material. This can be seen from Case IV of Fig. \ref{fig:cr_cases}, which shows that the interior of the bubble is covered by green curve (a tracer of wind material). In this case, the shock evolution is similar to Case II, although the shock energetics are different. For example, the stellar wind of a star can form a wind bubble. However, the outer shock of the bubble is weak because the ambient medium is hot. It also makes the bubble size smaller than in Case II. In contrast, the WTS can be as strong as in Case II. We have calculated the shock energetics and found that the shock energy is mostly encountered by the wind material, similar to Case III.

\subsection{Shock enegertics} \label{subsec:shockeng}
Let us summarise our findings here regarding the shock energetics, before moving on to the estimate of Neon isotope ratio:

\begin{itemize}
\item {\bf Isolated SN:}\\
In Case I, i.e., SN explosion in a uniform medium, the SN ejecta is confined inside a blast wave, and therefore can be accelerated by the reverse shock. We have shown that the reverse shock energy is $\sim$ ten times smaller than that of the forward shock (see bottom right panel of Fig. \ref{fig:c1}). Therefore, acceleration of SN ejecta is energetically {\it not} preferred in this case. We expect ISM nuclei to be accelerated more efficiently compared to SN nuclei by a ratio of $10:1$.  


In Case II (SN in the wind of a progenitor star), WTS can accelerate the wind material before the SN. When SN occurs, a blast wave moves through the free wind region for a few $100$ yr. Eventually, the forward shock of the blast wave collides with the WTS. During this phase, the reverse shock of the blast wave becomes strong (see left panel of Fig. \ref{fig:c2shockpf}) and finally reaches the center. In contrast, the forward shock of the blast wave moves through the hot bubble and hits the wind-driven shocked ISM shell. As the temperature of the bubble is $\sim 10^{7}$ K, the SN forward shock is weak compared to the reverse shock (see right panel of Fig. \ref{fig:c2shockpf}). We have estimated the shock energy encountered by the wind material and SN ejecta, and found that both are comparable. Therefore, in realistic calculations, acceleration of SN ejecta should also be considered in addition to the acceleration of the wind material. 
\item {\bf Star cluster:}\\
In this case, depending on the compactness of the star cluster, stellar winds from the stars can form a coherent WTS. This WTS can accelerate wind material for $\sim 3$ Myr before any SN. When the first SN occurs in the cluster, the forward shock of the blast wave moves outward by sweeping up the wind material. A reverse shock develops and sweeps up SN ejecta as well as the winds of remaining stars. When the SN forward shock collides with the WTS, it becomes weak. After that, it travels through the hot interior and the shock may disappear depending on the size of the bubble. Unlike Case - II, the SN reverse shock cannot reach to the center because the remaining stars continuously push it outward. This process brings the SN ejecta near the CD of the SB. After $\sim 10^{3}-10^{4}$ yr (depending on the compactness, see Figs. \ref{fig:c4} and \ref{fig:c3}), the SN reverse shock becomes weak and the winds of remaining stars again form a WTS. This process continues as long as the wind injection dominates over the SN event (until $\sim 10$ Myr). After that, the wind becomes weak and the SNe start controlling the evolution of the SB. SNe shocks continue to accelerate both wind material and SN ejecta. In the global energy budget, we have found that the shock energy deposited into wind material is larger than that into SN ejecta by a factor of $\sim 2-6$, depending on the compactness of the cluster:
\end{itemize}
\begin{figure}
\centering
\includegraphics[height= 3.2in,width=3.4in]{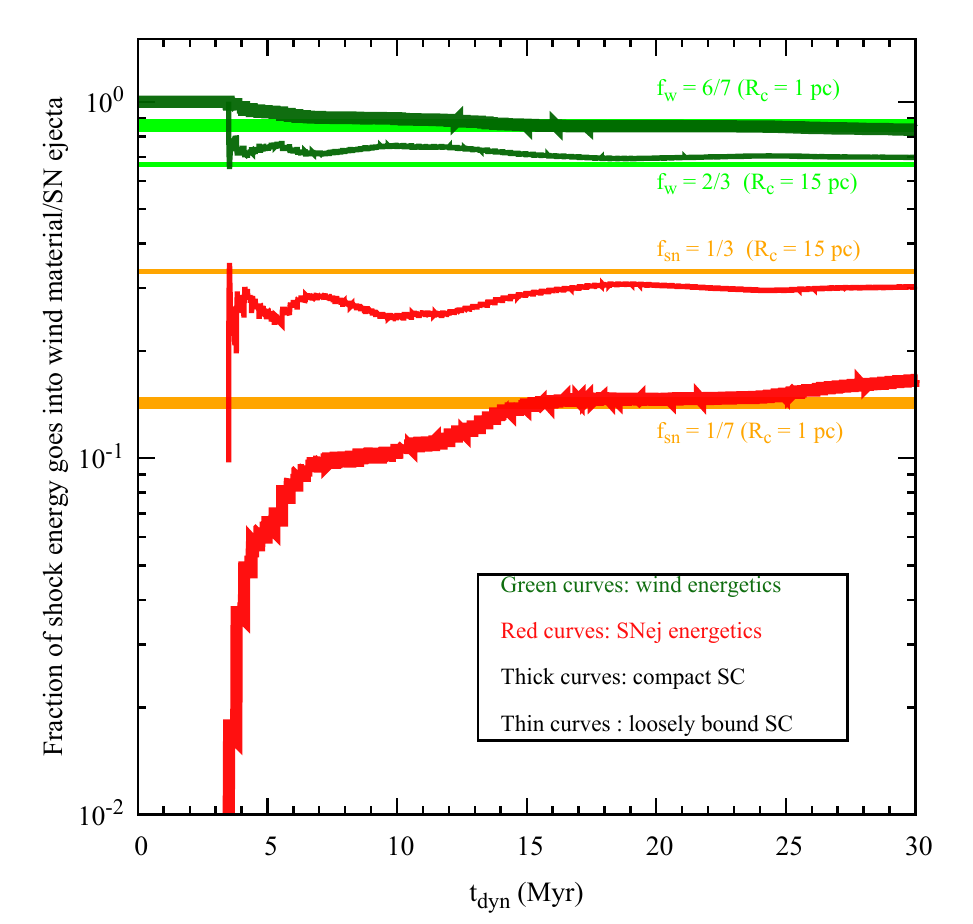}
\caption{Shock energetics of wind material and SN ejecta. The vertical axis shows the fraction of shock energy going into wind material (green) and SN ejecta (red) (Eq. \ref{eq:efrac}). Thick curves represent the case of compact star clusters  ($R_{\rm c}=1$ pc, i.e., Case III) and thin curves represent loosely bound star clusters ($R_{\rm c}=15$ pc, i.e., Case IV). These curves show that the shock energy encountered by wind material is larger than that by SN ejecta.}
\label{fig:efrac}
\end{figure}
\begin{figure*}
\centering
\includegraphics[height= 2.8in,width=6.5in]{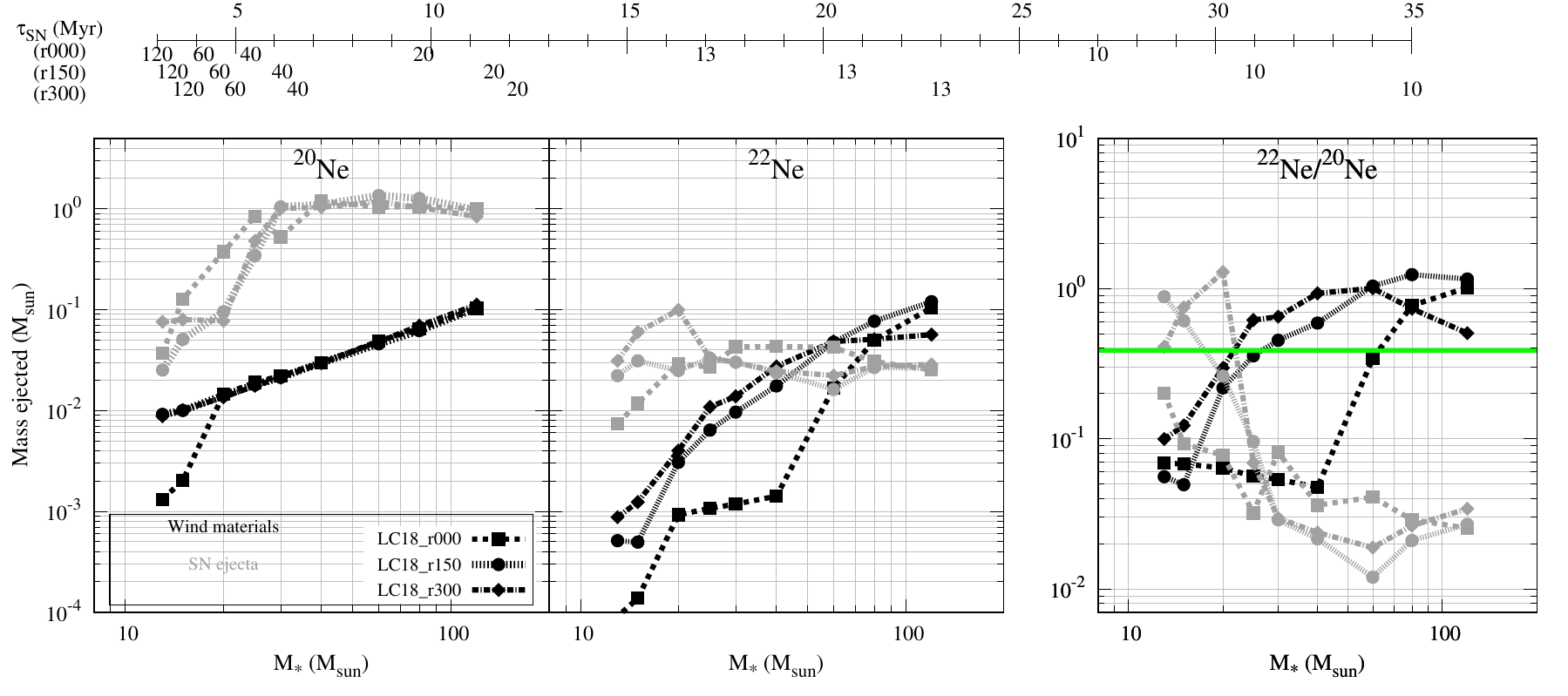}
\caption{Stellar wind (black) and SN (grey) yields for different stellar masses. Three line styles: dashed, dotted, and dash-dotted represent three stellar models where initial rotational speed are $0,\, 150$, and $300\, {\rm km\, s^{-1}}$ receptively. The top horizontal bar shows the epoch of SN explosion for some selective stellar masses ($M_{\rm *}$ in the unit of $M_{\rm \odot}$) for three different initial rotational speeds (e.g., the label $r300$ represents the rotational speed $300\,{\rm km\, s^{-1}}$). In rightmost panel, the green line shows the observed ${\rm ^{22}Ne/^{20}Ne}$ in GCRs.}
\label{fig:inputNe}
\centering
\includegraphics[height= 1.in,width=6.5in]{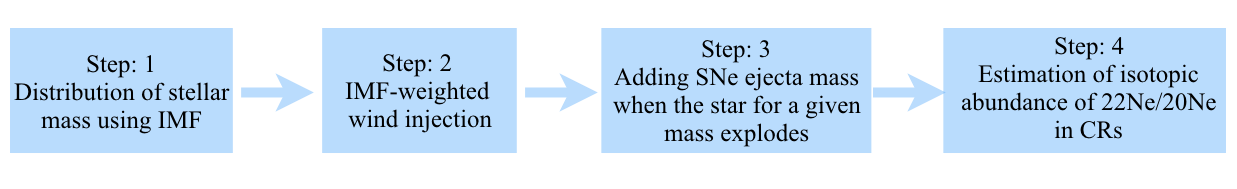}
\caption{Various steps to estimate ${\rm ^{22}Ne/^{20}Ne}$ ratio in CRs accelerated in star clusters.}
\label{fig:algo_yields}
\end{figure*}
 
Next we define two parameters to represent the fraction of shock energy processed in the wind material and SN ejecta in clusters, 
\begin{eqnarray} \label{eq:efrac}
f_{\rm w} =  \frac{E_{\rm w}}{E_{\rm w}+E_{\rm sn}} \ {\rm and} \ 
f_{\rm sn}  =   \frac{E_{\rm sn}}{E_{\rm w}+E_{\rm sn}} \ ,
\end{eqnarray}
where $E_{\rm w}$ and $E_{\rm sn}$ denote the total shock energy processed by wind material and SN ejecta respectively. These factors are shown in Fig. \ref{fig:efrac} as a function of dynamical time
\footnote{In case of isolated SN (Case I), the fraction of shock energy going into SN ejecta is $5\%$ and to ISM material is $95\%$ (see the bottom right panel of Fig. \ref{fig:c1}). For Case II of SN in massive stellar wind, these fractions become $\sim 70\%$ and $\sim 5\%$ respectively, and the rest $\sim 25\%$ goes into wind material (see the bottom right panel of Fig. \ref{fig:c2}).}.

This figure shows that $f_{\rm w} \approx 2/3  - 6/7$ and $f_{\rm sn} \approx 1/3 - 1/7$. At early times ($\lesssim 10$ Myr), the difference is larger than a factor of $10$ in case of a compact star cluster. {\it Therefore, in star clusters, the acceleration of the wind material is energetically preferred than the SN ejecta.}\footnote{Also see Appendix \ref{app:timedep}, which shows that the shock energy deposited into wind material is larger than that into SN ejecta by a factor of $\sim 2 - 6$ even when one considers a time-dependent wind model.}

   

\subsection{${\rm ^{22}Ne/^{20}Ne}$ in CRs} \label{sec:C}
Next we use our results of shock energetics to determine the isotope ratios in CRs, using the stellar evolutionary model of \citet{Limongi2018}. The yields of various elements in these models depend on initial metallicities, rotational velocities, and black hole cut off masses. Our estimates are robust to the variations within the uncertainties of these models. The abundances of ${\rm ^{20}Ne}$ and ${\rm ^{22}Ne}$ in wind and SN ejecta are shown in Fig. \ref{fig:inputNe}. In the rightmost panel, the curves indicate that ${\rm ^{22}Ne/^{20}Ne}$ ratio is large in the winds of massive stars (and to some extent, the SN ejecta of $\lesssim 20 M_\odot$ progenitors). The green line shows the corresponding ratio in Galactic CRs. Note that there are some difference in the ratios of ${\rm ^{22}Ne/^{20}Ne}$ between rotating and non-rotating models.

\begin{figure*}
\centering
\includegraphics[height= 2.5in,width=6.5in]{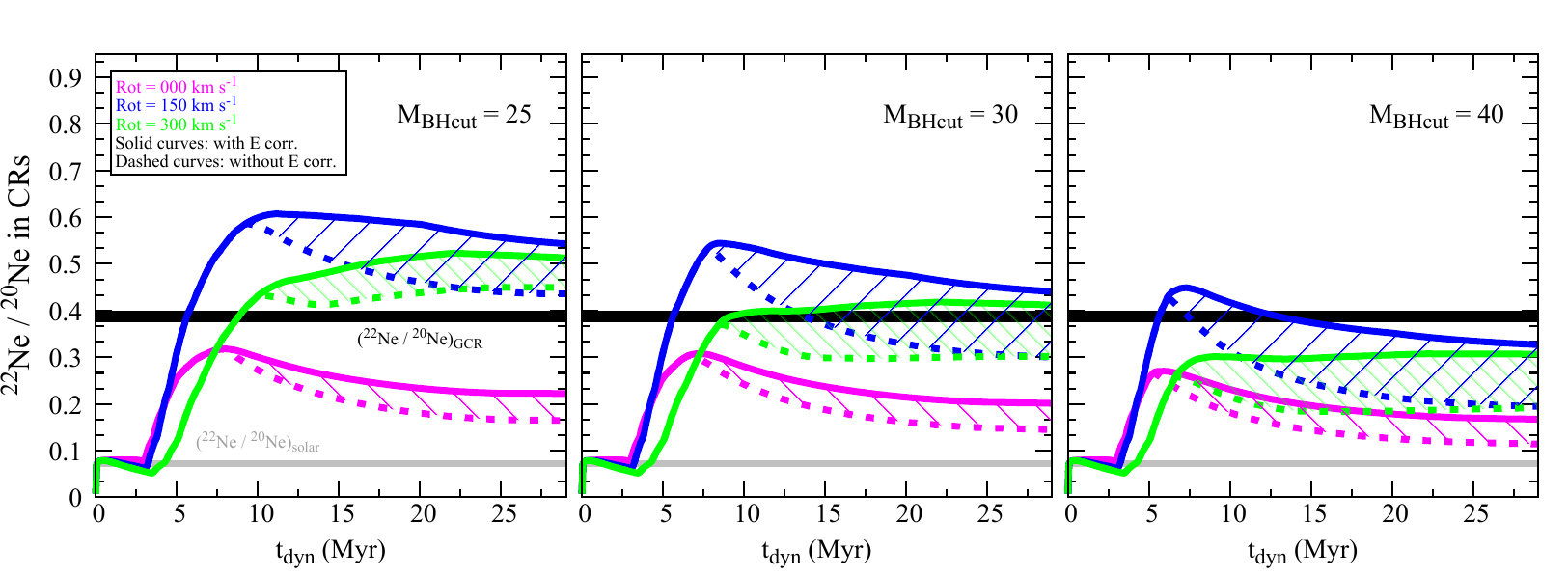}
\caption{Time evolution of ${\rm ^{22}Ne/^{20}Ne}$ in CRs from star clusters. Three panels show the result for models with different black hole cut-off mass ($M_{\rm BHcut}=25,30,$ and $40\, M_{\rm \odot}$). Three colours: magenta, blue, and green represent different stellar models with initial rotation speed $0, 150$ and $300\,{\rm km\,s^{-1}}$ respectively. Black and grey solid lines display the  ${\rm ^{22}Ne/^{20}Ne}$ ratio observed\citep{Binns2008} in GCR and in the solar wind respectively. Dashed curves show the  ${\rm ^{22}Ne/^{20}Ne}$ ratio without energy correction, i.e., acceleration of both wind material and SN ejecta are considered to be equally probable. Solid curves show the same ratio, after taking the energy weightage: $6/7$ and $1/7$ for wind material and SN ejecta respectively (see Figure \ref{fig:efrac}). Solid curves show that after energy correction, the time dependence of ${\rm ^{22}Ne/^{20}Ne}$ becomes weak and the values can be reconciled with observations.}
\label{fig:comNe}
\end{figure*}
Denoting the masses of elements ${\rm ^{22}Ne}$ and ${\rm ^{20}Ne}$ by ${\rm m^{^{22}Ne}}$ and ${\rm m^{^{20}Ne}}$ respectively, we estimate the isotopic ratio of ${\rm ^{22}Ne/^{20}Ne}$ as follows. Various steps of the calculation are summarised in Fig. \ref{fig:algo_yields}. Since the number of CRs accelerated is expected to be proportional to the energy crossing the shock, we calculate energy weighted wind and SN yields making use of shock energy deposited in wind material and SN ejecta, as calculated in \S \ref{subsec:shockeng}. The instantaneous isotopic ratio of ${\rm ^{22}Ne/^{20}Ne}$  in SB is
\begin{equation} \label{eq:Ne_accu}
\left(\frac{{m}^{^{22}Ne}}{{m}^{^{20}Ne}}\right) = \frac{f_{\rm w}\,\left \{\displaystyle \int_{t'=0}^t dt'\left(\displaystyle \sum_{i=1}^{(\rm N_{\rm *}- SN)} \dot{m}^{^{22}Ne}_{\rm w}(t') \right)\right \}+ f_{\rm sn}\,\left \{ \displaystyle \sum_{i=1}^{\rm SN} m^{^{22}Ne}_{\rm sn} \right \}}{f_{\rm w}\,\left \{\displaystyle \int_{t'=0}^t dt'\left( \displaystyle\sum_{i=1}^{(\rm N_{\rm *}- SN)} \dot{m}^{^{20}Ne}_{\rm w}(t') \right)\right \}+ f_{\rm sn }\,\left \{\displaystyle \sum_{i=1}^{\rm SN} m^{^{20}Ne}_{\rm sn}\right\}}
\end{equation}
The time average ratio of ${\rm ^{22}Ne/^{20}Ne}$ that is accelerated up to time $t$ in SB (\citealt{Prantzos2012}) is obtained using
\begin{equation}\label{eq:Ne_result}
\left(\frac{{^{22}Ne}}{{^{20}Ne}}\right)_{\rm CR} = \frac{1}{t}\int_{t'=0}^t dt' \left(\frac{{m}^{^{22}Ne}}{{m}^{^{20}Ne}}\right)  
\end{equation}
These ratios for different stellar parameters are shown in Fig. \ref{fig:comNe}.

Figure \ref{fig:comNe} shows that the ratio of $^{22}{\rm Ne}/^{20}{\rm Ne}$ is dominated by the abundance ratio in the stellar wind material. Initially the ratio is small because the wind of massive stars is dominated by $^{20}{\rm Ne}$ except in the pre-supernova wind of stars with mass $\ge 20\hbox{--}60$ M$_\odot$, which begin to appear only after $\sim 3$ Myr. Therefore, $^{20}{\rm Ne}$ produced in SN ejecta and winds of lower mass stars remain sub-dominant in CRs. After $\sim 10$ Myr, the cluster is left with stars ($\le 20$ M$_\odot$) that contribute little to ${\rm ^{22} Ne}$ and $^{20}{\rm Ne}$, and the isotope ratio remains practically frozen at the value  attained by this time. Although there is a clear enhancement of ${\rm ^{22} Ne}/^{20}{\rm Ne}$ has been observed in our analysis, it is worth mentiong that the stellar yields depend on the assumption of rotational speed of massive stars. We also note that since the acceleration of ejecta material is energetically less efficient, other isotope ratios such as ${\rm ^{59}Ni/^{59}Co}$ are likely to be small in GCR, as observed \citep{Wiedenbeck1999}.

\section{Discussions}
We have therefore been able to calculate the relative contribution of WTS and SNe shocks in star clusters, and their role in enhancing the Neon isotope ratio to the observed label. This match with the observed ratio encourages us to speculate on the implications of our calculation.

\subsection{Common platform for WTS and SNe shocks}
To begin with, our calculations make use of both SNe shocks and WTSs in star clusters. It may appear that isolated SNRs, which are considered as the standard sites of CR acceleration, have no relation with star clusters. However, star clusters provide the necessary ingredients for an integrated picture of both sources (SN shocks and wind termination shocks), considering the fact that massive OB stars form in dense, compact (of pc scale) clusters (\citealt{PortegiesZwart2010}; \citealt{Pfalzner2009}). Although massive stars are sometimes found outside clusters, as in LMC (\citealt{Sana2013}), they can be considered to be `slow runaways' from massive clusters (\citealt{Banerjee2012}; \citealt{Lucas2018}). Although diffuse or `leaky' clusters (\citealt{Pfalzner2009}) have sizes of about an order of magnitude larger than compact clusters (e.g., Cyg OB2), they can be thought of as assemblies of compact clusters, given that they have substantial substructures (\citealt{Wright2014}). Finally, if the initial cluster is small enough, it can be dissolved (or nearly dissolved) due to gas dispersion and mass loss during SNe (\citealt{Brinkmann2017}, \citealt{Shukirgaliyev2017}). 

Observations therefore indicate that all OB stars, including the progenitor isolated SNRs, can be considered to have formed in massive compact clusters.
\subsection{Number statistics of SNe and massive clusters} \label{sec:D}
Based on above discussions, we can therefore argue that the massive progenitors of apparently isolated SNRs can be considered as belonging to  clusters in which the number of massive (OB) stars is $\lesssim 2$. The cluster luminosity function \citep{Williams1997} is observed to be  $dN/dN_{\rm OB}\propto N_{\rm OB}^{-2}$, where $N_{\rm OB}$ is the number of OB stars in a cluster. This implies that roughly half the clusters would produce an apparently isolated (core collapse) SNR, since (for a lower limit of a massive star cluster being that with a single OB star)
\begin{equation}
{\int_1 ^2 N_{\rm OB} ^{-2} dN_{\rm OB} \over \int_1 N_{\rm OB} ^{-2} dN_{\rm OB}}\approx {1 \over 2}
\end{equation}
which is almost independent of the upper limit of total number of OB stars in a clusters which is roughy $\sim 7000$ (\citealt{Mckee1997}).

In comparison, $\gamma$-ray bright clusters ($\gamma$-ray luminosity $\gtrsim 10^{35}\, {\rm erg \,s^{-1}}$) have typically $\sim 50\hbox{--}100$ OB stars \citep{Gupta2018b}. This mass function implies a ratio of core collapse SNRs to WTS gamma-ray sources is  
\begin{equation} 
{\int_1 ^2 N_{\rm OB}^{-2} dN_{\rm OB} \over \int_{50\hbox{--}100} N_{\rm OB} ^{-2} dN_{\rm OB}} \approx {1\over 2} (50 \hbox{--}100)= (25\hbox{--}50)\,.
\end{equation} 
Observations show that in Sbc galaxies (Milky-Way-type) the ratio of thermonuclear to core collapse SNe is  $\sim 1/3$ \citep{Mannucci2005}. Adding thermonuclear SNRs, this  indicates a ratio of SNRs to WTS $\gamma$-ray sources in our Galaxy of order $33\hbox{--}65$.

We can check this argument by counting the SNRs and $\gamma$-ray bright clusters in a given volume. The size of the sampling volume is hard to decide, since $\gamma$-ray bright clusters are rare: there are only 3 $\gamma$-ray bright clusters near us, Cygnus at $\approx 1.4$ kpc, Westerlund 1 at $\approx 4$ kpc \citep{Aharonian2019}, and Westerlund 2 at $\approx 5$ kpc \citep{Yang2018}. Therefore, it makes sense to use all these three and use a sphere of radius $5$ kpc around us. Counting the number of SNRs from the available catalogue, we find $124$ of them within a distance of $5$ kpc (listed in Table \ref{tab:SNRs} of Appendix \ref{app:SNRtab}), so that the ratio $124:3$ falls in the ballpark of the above estimate.
  
 This match can be regarded  as a consistency check for our proposed scenario, and this   implies that isolated SNRs can indeed be considered as part of the phenomena of star clusters. Therefore, the two types of CR acceleration sites can indeed be put on a common platform.

\subsection{Contributions of WTS and SN shocks in total energy budget of GCRs}\label{subsec:totengbudget}
\begin{figure}
\centering
\includegraphics[height= 3.2in,width=3.4in]{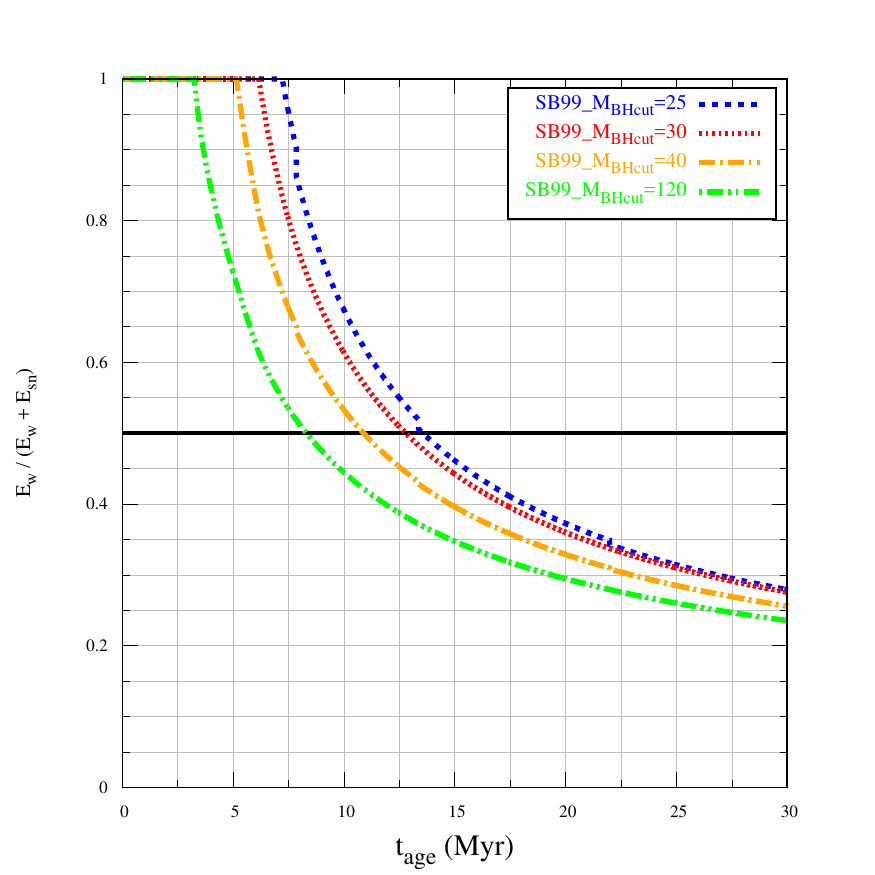}
\caption{
Contribution to stellar winds in the total mechanical energy output of a typical star cluster until a given epoch. All curves are obtained using the Starburst99 stellar synthesis code \citep{Leitherer1999} for the Kroupa Initial Mass Function \citep{Kroupa2002}. Each curve represents a model with different black hole mass, $M_{\rm BHcut}$ (blue - $25\, M_{\rm \odot}$, red - $30\, M_{\rm \odot}$, yellow - $40\, M_{\rm \odot}$, and green - $120\, M_{\rm \odot}$). The figure shows that wind mechanical energy dominates over SN energy before $\gtrsim 11\pm 3$ Myr.}
\label{fig:comp_en}
\end{figure}

The next issue to consider is the energy budget of Galactic CRs from these two sites. It can be shown that these two classes of sources are energetically comparable (as pointed out in $1980$s, see e.g., \citealt{Abbott1981}), WTS being dominant in young phase of clusters. Figure \ref{fig:comp_en} plots the evolution of the contribution of stellar winds to the total mechanical energy of a star cluster, using Starburst99 \citep{Leitherer1999}, and Kroupa initial mass function \citep{Kroupa2002}, for different assumptions of black hole mass cutoff. The curves in the figure show that the fraction of the total shock energy in WTS  for a massive cluster is always $\geq 0.25$, and is larger than $0.5$ for clusters younger than $\approx 10$ Myr, thereafter decreasing with time because of the increasing number of SNe.  
Therefore, {\it at least a quarter, if not a larger fraction, of the Galactic CRs can be ascribed to WTSs in star clusters}.

For typical parameters the average mechanical power in stellar winds within the solar circle \citep{Reed2005} is $L_w \sim 10^{36} \, {\rm erg} \, {\rm s}^{-1} \, N_{\rm OB} \approx 2 \times 10^{41} \, {\rm erg} \, {\rm s}^{-1}$ consistent with recent estimates \citep{Seo2018}. If this represents a quarter of the total mechanical energy in SNS and WTS, then this implies a mechanical energy budget for SNSs of $6 \times 10^{41}$ erg s$^{-1}$, which in turn corresponds to a SN rate of $\sim 2$ per century, consistent with the observed rate of $2$ per century \citep{Diehl2006}.
Incidentally, with a typical efficiency of CR acceleration $\sim 0.1$, the total mechanical luminosity of $8 \times 10^{41}$ erg s$^{-1}$ is comparable to the inferred Galactic CR (GCR) luminosity \citep{Strong2010} of $\sim 8 \times 10^{40}$ erg s$^{-1}$.

Therefore, the scenario of CR acceleration in WTS and SNSs is consistent with respect to the energy budget.

\subsection{Maximum energy of CRs}\label{subsec:maxcreng}
 The importance of WTS as CR sources becomes crucial at high energies \citep{Cesarsky1983}, since the observed $\gamma$-ray spectra of clusters are flatter than that of SNRs \citep{Aharonian2019}. The maximum CR energy depends  on the extent of the accelerating region, in this case, the width of the shocked wind region, which  separates the contact discontinuity and WTS. 
The distance of the
WTS is roughly given by 
\begin{equation}
R_{ts}\approx 25 \, {\rm pc} \, n_{1}^{-3/10} \, \dot{M}_{-4} ^{1/2} \, L_{\rm w,38}^{-1/5} \, v_{\rm w,2000}^{1/2} \, t_{\rm 3M}^{2/5}\,,
\end{equation}
 where $n_{1}$ is the ambient particle density in units of $10$ cm$^{-3}$, $\dot{M}_{-4}$ is mass loss rate in units of $10^{-4}$ M$_\odot$ yr$^{-1}$,  $v_{w,2000}$ is wind speed in units of $2000$ km s$^{-1}$, $L_{\rm w,38}$ is wind mechanical power in units of $10^{38}$ erg s$^{-1}$, $t_{\rm 3M}$ is time in 3 Myr unit. The distance of the contact discontinuity is $ (\eta L_w t^3/\rho )^{1/5}$, where $\eta\sim 0.2$ takes into account energy lost in radiative cooling (see e.g., \citealt{Sharma2014}). For the same fiducial parameters, this distance is given by $\sim 62$ pc. Therefore, for a $10^4$ M$_\odot$ cluster, the extent of the shocked wind region at $\sim 3$ Myr (when WTS dominates) is $\sim 40$ pc. This estimate is also borne out with our 1-D simulation (see e.g., Figure \ref{fig:cr_cases} for compact star cluster, i.e., Case III). 

According to the Hillas criterion \citep{Hillas1984}, the accelerating region should be larger than $2r_L/\beta$ where $r_L$ is the Larmor radius of a particle, which leads to a maximum energy 
 \begin{equation}\label{eq:Emax}
 E \le {1\over 2} 10^{15} \, {\rm eV} \, L_{\rm pc} B_{\rm \mu G} \, Z \, \beta \,,
 \end{equation} 
 where $\beta$ is the ratio of shock speed to that of light, and $L_{\rm pc}$ is the extent of the accelerating region in parsec.\footnote{\citet{Biermann2018} argued that the `$\beta$' term in Eq. (\ref{eq:Emax}) may be absent, depending on the magnetic field configuration, see e.g., \citet{Jokipii1987}} With a $10\mu$G field, and wind velocity $\sim 2000\, {\rm km\,s^{-1}}$, this implies a maximum energy \citep{Hillas1984} of $\sim 1.5 \, Z$ PeV, much larger than in isolated SNR. We note that \cite{Voelk1988} had argued that SNe shocks expanding in a wind bubble could circumvent the problems with isolated SNRs, but our prediction is independent of SNe event in a star cluster. We also note that LOFAR observations show PeV CRs to be enriched in low-Z nuclei \citep{Buitink2016}, consistent with their being accelerated from CNO enriched wind material in SBs \citep{Thoudam2016}. It is also possible that the hard component of CRs due to WTS is the one inferred to be present in molecular clouds \citep{Boer2017}. Therefore WTSs are potential PeVatrons and they should be studied in detail in this regard.

\subsection{Decoupling of grammage from ISM}\label{subsec:grammage}
Our proposed scenario can also provide the astrophysical framework for the phenomenologically motivated  models of CR propagation in which the grammage traversed by CRs are mostly near the source (\citealt{Cowsik2016,Eichler2017,Blasi2009,Biermann2018}). In these models  the diffusion property of CR particles is assumed to be different inside the `cocoons' surrounding the CR acceleration sites than elsewhere in the ISM in order to alleviate problems associated with secondary production. 


The scenario of CR acceleration in SBs satisfies the basic premises of these models, since most of the CR collisions occur in the outer shell of shocked ISM, whereas CRs are advected by the wind in the inner regions. The grammage suffered by CRs in the shocked ISM region is $\sim 10$ g cm$^{-2}$, comparable to the total grammage of Galactic CR \citep{Gabici2019}, since the typical residence time is $t_{\rm r}\sim 1$ Myr ($\sim \kappa/v^2$, for a diffusion coefficient $\kappa\sim 10^{27}$ cm$^2$ s$^{-1}$ and $v\sim 50$ km s$^{-1}$, the typical outer shock speed), and typical density in this region is $\sim 10$ cm$^{-3}$. 

Therefore, CR grammage can be decoupled from Galactic residence time, alleviating a number of outstanding problems \citep{Butt2009,Gabici2019}. Instead of listing them all, we mention  one of the problems here, that of  the observed scaling of light elements with metallicity in halo stars \citep{Parizot1999}. 
 In the case of isolated SNRs (which accelerates ISM particles), the Li/Be/B abundance is expected to scale as $Z^2$, since both CR and the target gas share the same metallicity. In the case of WTS, the metallicity of CRs is independent of that of the ambient ISM, the abundance of spallation products should scale as $Z$, as observed \citep{Parizot1999}.

\section{Conclusions}
With 1-D numerical hydrodynamical simulation of stellar winds in star clusters, we have studied the relative importance of SNe shocks and wind termination shocks (WTSs) in clusters, both compact and massive, and also in the case of SNe shocks running into stellar winds of the progenitor star. Our findings are as follows:
\begin{enumerate}
\item WTSs process more than $1/4$ of the total mechanical power in a star cluster, and this fraction rises to $\ge 0.5$ for young ($\le 10$ Myr) clusters (Figure \ref{fig:comp_en}). Therefore a significant fraction of Galactic CRs are accelerated in these shocks in massive compact star
clusters.
\item A large fraction ($\approx 2/3\hbox{--}6/7$) of the total energy processed by WTS and SNe shocks goes to accelerate the stellar wind material, enriched in $^{22}{\rm Ne}$ (Figure \ref{fig:efrac}). Using this ratio as an energy weightage for the stellar wind and SN ejecta, we show that the time averaged ratio of $^{22}{\rm Ne}/^{20}{\rm Ne}$ matches the observed value.
\item We also show that the scenario often quoted in literature for explaining the Neon ratio, namely, by SNe shocks in a stellar wind of progenitor star, is problematic because we find that the reverse shock in this case is as efficient as the forward shock, and it would accelerate SN ejects (rich in $^{20}{\rm Ne}$) posing a problem for the isotope ratio.
\item The combined effect of WTSs and SNe shocks in star clusters can explain the observed Neon isotope ratio (Figure \ref{fig:comNe}), unlike previous approaches of either SNe shocks (in stellar winds) or only stellar winds (e.g., \citealt{Binns2008}).
\item We take this approach forward to suggest that these two sources can be brought under a same umbrella of CR acceleration in SBs, since SNe and stellar wind are both related to massive stars and they form in clusters. Isolated SNRs can be considered as representing the lower end of the star cluster mass function.
\item Using the luminosity function of OB associations, we show that the expected ratio of isolated SNRs to $\gamma$-ray bright star clusters matches observed numbers (section \ref{sec:D}).
\item We argue that the accelerating region in the case of WTSs is large enough to explain CRs in PeV range (section \ref{subsec:maxcreng}), thus making WTSs an important complementary source of CRs.
\item The fact that most of the CR interactions occur in the SB, with a grammage that corresponds to the observed grammage of GCRs (section \ref{subsec:grammage}), decouples the CR grammage from the Galactic residence time and helps solve other long standing problems with the standard paradigm of SNRs as CR acceleration sites.
\end{enumerate}

The integrated picture of CR acceleration in SBs inflated by clustered WTS and SN presented here can solve these problems as well as explain $^{22}{\rm Ne}/^{20}{\rm Ne}$ ratio, while being consistent with the number statistics of SNRs and WTS sources.

\section*{Acknowledgements}
We thank S. Banerjee, P. L. Biermann, P. Chandra and N. Prantzos for useful discussions. PS thanks the Humboldt Foundation for supporting his sabbatical at MPA. PS acknowledges the Department of Science and Technology for a Swarnajayanti Fellowship (DST/SJF/PSA-03/2016-17). SG thanks CSIR India for the SPM Fellowship. 



\bibliographystyle{mnras}
\bibliography{references}


\onecolumn 
\appendix
\section{3-D set-up} \label{subsec:num3d}
Here we present the set-up of our 3-D simulations discussed in section \ref{subsec:formation_wts}. We simulate two star clusters with two different core radii $R_{\rm c}$. Both clusters have twelve massive stars (i.e., $N_{\rm OB}=12$), which corresponds to a cluster of stellar mass $10^{3}\, M_{\rm \odot}$ (for a standard Kroupa initial mass function (IMF) with lower- and upper- cutoff masses $0.1\,M_{\rm \odot}$ and $120\,M_{\rm \odot}$; \citealt{Kroupa2002}). We have solved the standard Euler equations in $3$-D Cartesian geometry using the {\bf\textsf{\small PLUTO}} code (\citealt{Mignone2007}), where the HLLC Riemann solver is used (CFL number $0.3$) and the computational box extends from $-60$ pc to $+60$ pc along $x, y$, and $z$ directions. At $t=0$, we assume that ambient density is uniform with $\rho = 50\, m{\rm_{\rm H}\,cm^{-3}}$, and pressure is $10^{-12}\, {\rm dyne\,cm^{-2}}$. Radiative cooling has been included both runs. Locations of the stars are randomly chosen by making use of a Gaussian random number generator, where we have used two sets of mean deviations, $\sigma_{\rm x,y,z} = 0.5$ pc and $\sigma_{\rm x,y,z} = 5$ pc, to represent compact and loosely bound star clusters respectively. For each star, we have assumed $\dot{M}=8\times 10^{-7}\, M{\rm _{\odot}\,yr^{-1}}$ and $L_{\rm w}=10^{36}\, {\rm erg\,s^{-1}}$, which are injected in small spherical regions of radius $\delta r_{\rm inj} =0.2$ pc (for details see section 4.2 in \citealt{Gupta2018b}). 
\section{Time dependent luminosity model} \label{app:timedep}
To estimate the fraction of shock energy encountered by wind material and SNe ejecta, in the previous sections we have considered a time-independent wind model, where it has been assumed the wind mechanical luminosity of each star in a cluster is $10^{36}\, {\rm erg\,s^{-1}}$ until it explodes as SN. Here we investigate the effect of the time-dependent wind luminosity in our calculation.

\begin{figure}
\begin{minipage}[t]{0.63 \textwidth}
\centering
\includegraphics[height= 2.4in,width=4.5in]{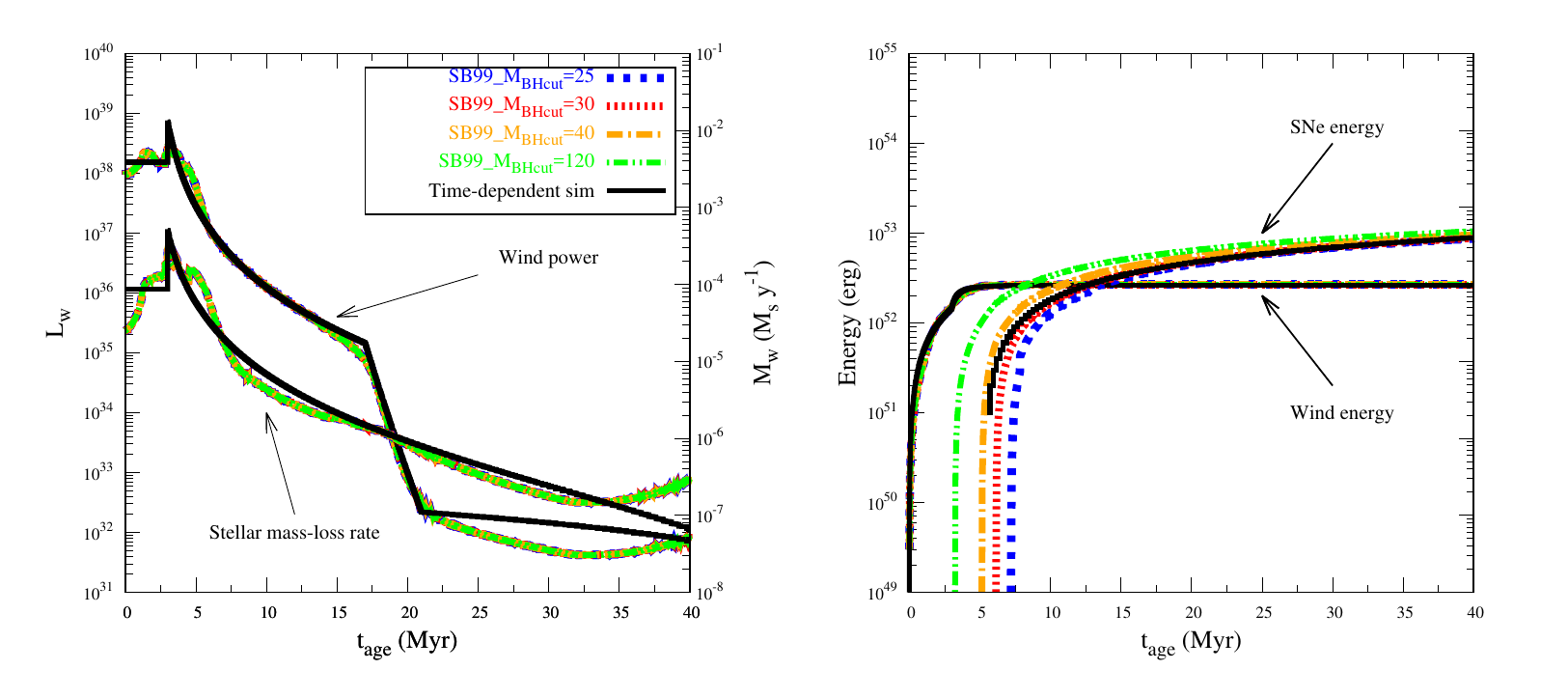}
\caption{Time dependent wind model. [Left panel] Time evolution of stellar wind luminosity and mass-loss rate of a cluster of mass $10^{4}\, M_{\rm \odot}$ obtained using the Starburst99 stellar synthesis code (\citealt{Leitherer1999}). Each curve represents a model with different black hole cut-off mass, $M_{\rm BHcut}$ (blue - $25\, M_{\rm \odot}$, red - $30\, M_{\rm \odot}$, yellow - $40\, M_{\rm \odot}$, and green - $120\, M_{\rm \odot}$). The figure shows that stellar wind luminosity decreases and becomes almost negligible after $\gtrsim 15$ Myr. [Right panel] Comparison between wind mechanical energy and SN energy in a cluster. Different colours show that the time when SN energy dominates over wind energy depends on $M_{\rm BHcut}$ and it occurs after $\approx 11\pm3$ Myr. In both panels, the black curves represent an injection model used to show the effect of time-dependent wind model in our calculation.}
\label{fig:timedep}
\end{minipage}
\hspace{0.02\textwidth}
\begin{minipage}[t]{0.35 \textwidth}
\centering
\includegraphics[height= 2.5in,width=2.5in]{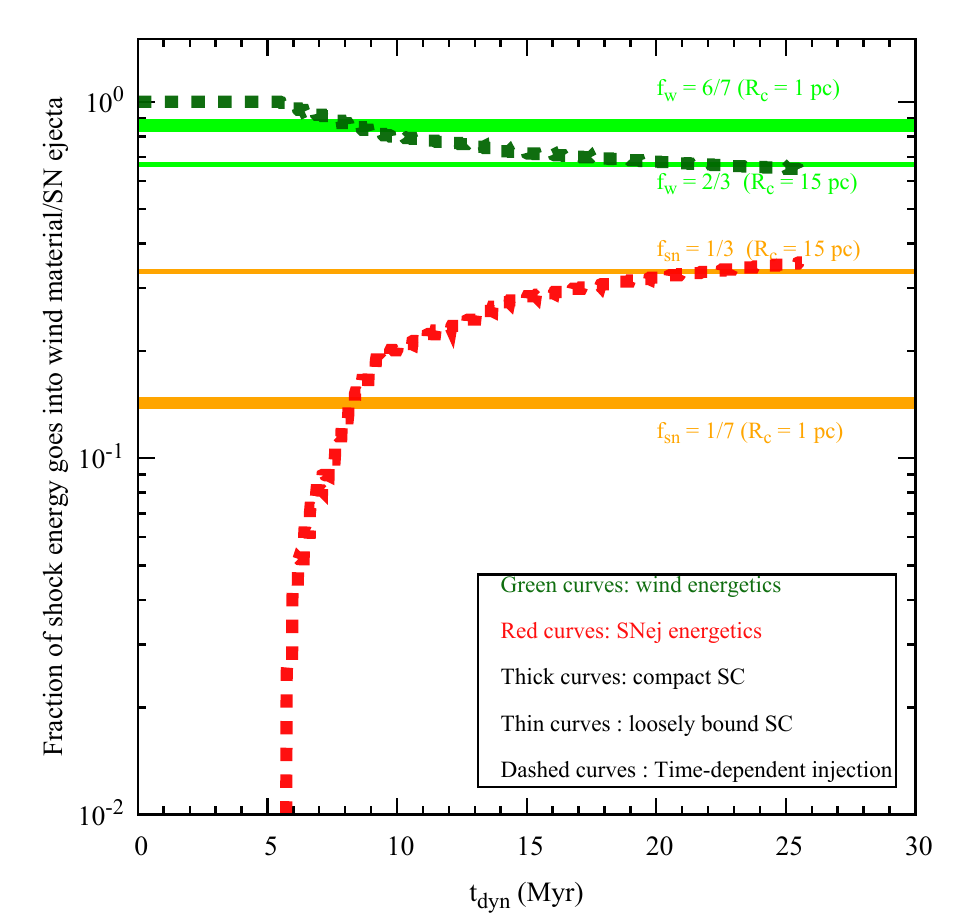}
\caption{Shock energetics of wind material (green) and SN ejecta (red) for a time dependent wind model (also see Fig. \ref{fig:efrac}). The dashed curves represent a time-dependent wind model for a compact star cluster where mass and energy have been injected according to the black curves in Fig. \ref{fig:timedep}. These curves confirm that the shock energy encountered by wind material is larger than that by SN ejecta.}\label{fig:efrac_dep}
\end{minipage}
\end{figure}

We use a stellar synthesis code {\bf\textsf{\small
Starburst99}} (\citealt{Leitherer1999}) to obtain the time evolution of wind luminosity, stellar mass-loss rate, and SNe energy of a cluster of mass $10^{4}\, M_{\rm \odot}$  for four different black hole cut-off masses: $25$, $30$, $40$, and $120\, M_{\rm \odot}$. The wind mechanical power and mass-loss rates are shown in the left panel of Fig. \ref{fig:timedep}. Figure shows that the wind mechanical power and mass-loss rates do not depend on the black hole cut-off mass, as expected. The right panel shows the total energy deposited by SNe and by winds in the cluster until a given epoch. The black curves represent fitted curves used in our model to inject stellar wind and SN energy. The fraction of shock energy processed in wind material and SNe ejecta are shown in Fig. \ref{fig:efrac_dep}. The figure shows that the fraction of shock energy processed in wind is larger than that in SN ejecta.

\section{SNRs} \label{app:SNRtab}
Table \ref{tab:SNRs} shows a list of $124$ SNRs that has been used in section \ref{sec:D} to discuss number statistics of SNe and massive clusters.
\begin{table*}
\centering
\caption{This table provides a list of SNRs which are located within $5$ kpc. [1] and [2] refer to
[1] \href{https://arxiv.org/abs/1409.0637}{https://arxiv.org/abs/1409.0637}; \href{https://www.mrao.cam.ac.uk/surveys/snrs/}{https://www.mrao.cam.ac.uk/surveys/snrs/}  and
[2] \href{https://hea-www.cfa.harvard.edu/ChandraSNR/snrcat\_gal.html}{https://hea-www.cfa.harvard.edu/ChandraSNR/snrcat\_gal.html}.}
\scriptsize
\begin{minipage}[b]{0.40\linewidth}
\centering
\vspace{0.5em}
\begin{tabular}{ c c c c}
  \hline\hline
  &   Galactic coordinates  & Distance & Ref.\\
  & $ (l,b)$ & (kpc) & \\
\hline
  1. &  $4.5,\ +6.8$ & $3.3-5.1$ & [1]\\
  2. &  $5.2,\ -2.6$ & $4.3-5.2$ & [1]\\
  3. & $5.7,\ -0.1$ & $2.9-3.2$ & [2]\\
  4. & $6.4,\ -0.1$ & $2.2$ & [1]\\
  5. & $7.7,\ -3.7$ & $3.2-6$ & [1]\\
  6. & $8.7,\ -0.1$ & $4.0$ & [1]\\
  7. & $11.0,\ -0.0$ & $5.0$ & [1]\\
  8. & $11.1,\ +0.1$ & $4.4$ & [1]\\
  9. & $13.3,\ -1.3$ & $2-4$ & [1]\\
  10. & $5.7,\ -0.1$ & $3.2$ & [1]\\
  11. & $7.5,\ -1.7$ & $1.7-2.0$ & [2]\\
  12. & $15.1,\ -1.6$ & $2.1$ & [1]\\
  13. & $15.4,\ +0.1$ & $4.8$ & [1]\\
  14. & $18.6,\ -0.2$ & $4.2-4.6$ & [1]\\
  15. & $18.9,\ -1.1$ & $4.2-4.6$ & [1]\\
  16. & $19.1,\ +0.2$ & $1-2$ & [1]\\
  17. & $21.5,\ -0.9$ & $4.7$ & [1]\\
  18. & $22.7,\ -0.2$ & $4.5-4.9$ & [1]\\
  19. & $23.3,\ -0.3$ & $4.6-5.0$ & [1]\\
  20. & $24.7,\ -0.6$ & $-4$ & [1]\\
  21. & $24.7,\ +0.6$ & $2-3.7$ & [1]\\
  22. & $28.8,\ +1.5$ & $3.8-4.0$ & [1]\\
  23. & $29.6,\ +0.1$ & $4.4-5.0$ & [1]\\
  24. & $29.7,\ -0.3$ & $5.0$ & [1]\\
  25. & $32.0,\ -4.9$ & $1.8$ & [1]\\
  26. & $32.1,\ -0.9$ & $4.6$ & [1]\\
  27. & $32.8,\ -0.1$ & $4.5-5.1$ & [1]\\
  28. & $34.7,\ -0.4$ & $2.1-3$ & [1]\\
  29. & $40.5,\ -0.5$ & $3.2$ & [1]\\
  30. & $42.0,\ -0.1$ & $3.1-3.9$ & [1]\\
  31. & $49.2,\ -0.7$ & $4.3$ & [1]\\
  32. & $53.6,\ -2.2$ & $2.8$ & [1]\\
  33. & $55.7,\ +3.4$ & $0.56$ & [1]\\
  34. & $57.2,\ +0.8$ & $3.5$ & [1]\\
  35. & $21.5,\ -0.9$ & $4.7$ & [2]\\
  36. & $21.9,\ -0.1$ & $4.3$ & [2]\\
  37. & $23.5,\ +0.1$ & $5$ & [2]\\
  38. & $26.6,\ -0.51$ & $1.3$ & [1]\\
  39. & $34.0,\ +20.3$ & $1.4$ & [1]\\
  40. & $59.2,\ -0.47$ & $2.5$ & [1]\\
  41. & $65.3,\ +5.7$ & $0.8 $ & [1]\\
  42. & $65.7,\ +1.2$ & $ 1.5$ & [1]\\
  43. & $65.8,\ -0.5$ & $ 1.9-2.7$ & [1]\\
  44. & $66.0,\ -0.0$ & $ 2.0-2.6$ & [1]\\
  45. & $67.6,\ +0.9$ & $ 1.8-2.2$ & [1]\\
  46. & $70.0,\ -21.5$ & $ \lesssim 3 $ & [1]\\
  47. & $73.9,\ +0.9$ & $0.5-4.0$ & [1]\\
  48. & $74.0,\ -8.5$ & $0.44 $ & [1]\\
  49. & $78.2,\ +2.1 $ & $ 1.7-2.6$ & [1]\\
  50. & $82.2,\ +5.3$ & $1.3 -3.2$ & [1]\\
  51. & $85.4,\ +0.7$ & $3.5 $ & [1]\\
  52. & $85.9,\ -0.6$ & $ 4.8$ & [1]\\
  53. & $89.0,\ +4.7$ & $0.8$ & [1]\\
  54. & $93.3,\ +6.9$ & $ 2.2$ & [1]\\
  55. & $93.7,\ -0.2$ & $1.5 $ & [1]\\
  56. & $10.9,\ -45.4$ & $ 0.25$ & [2]\\
  57. & $25.1,\ -2.3$ & $2.9 $ & [1]\\
  58. & $33.6,\ +0.1$ & $ 3.5-7.1$ & [1]\\
  59. & $39.7,\ -2.0$ & $ 3.5-6.5$ & [1]\\
  60. & $42.8,\ +0.6$ & $2.8-7.7$ & [1]\\
  61. & $69.0,\ +2.7$ & $ 1.5$ & [1]\\
  62. & $106.3,\ +2.7$ & $0.8-3.0$ & [1]\\
  63. & $126.2,\ +1.6$ & $4.5 $ & [1]\\
 \hline
\end{tabular}\\
\end{minipage}
\hspace{0.1\linewidth}
\begin{minipage}[b]{0.40\linewidth}
\centering
\begin{tabular}{ c c c c}
  \hline\hline
  &   Galactic coordinates  & Distance & Ref.\\
  & $ (l,b)$ & (kpc) & \\
\hline
64. & $152.4,\ -2.1$ & $ 1$ & [1]\\
65. & $284.3,\ -1.8$ & $ 2.9$ & [1]\\
66. & $299.2,\ -2.9$ & $5 $ & [1]\\
 67. & $96.0,\ +2.0 $ & $ 4$ & [1]\\
  68. & $108.2,\ -0.6$ & $ 3.2$ & [1]\\
  69. & $109.1,\ -1.0$ & $ 3.2$ & [1]\\
 70. & $111.7,\ -2.1$ & $3.3-3.7$ & [1]\\
  71. & $113.0,\ +0.2$ & $ 3.1$ & [1]\\
  72. & $114.3,\ +0.3 $ & $0.7$ & [1]\\
  73. & $116.5,\ +1.1$ & $1.6$ & [1]\\
  74. & $116.9,\ +0.2$ & $1.6 $ & [1]\\
  75. & $119.5,\ +10.2$ & $ 1.4$ & [1]\\
  76. & $120.1,\ +1.4 $ & $2.4 $ & [1]\\
  77. & $127.1,\ +0.5$ & $ 1.2-1.3$ & [1]\\
  78. & $130.7,\ +3.1$ & $2 $ & [1]\\
  79. & $132.7,\ +1.3$ & $2$ & [1]\\
  80. & $156.2,\ +5.7$ & $ 1.7$ & [1]\\
  81. & $160.9,\ +2.6$ & $<4$ & [1]\\
  82. & $166.0,\ +4.3$ & $ 4.5$ & [1]\\
  83. & $178.2,\ -4.2$ & $ 2.9$ & [1]\\
  84. & $180.0,\ -1.7$ & $ 0.36-0.88$ & [1]\\
  85. & $184.6,\ -5.8$ & $2 $ & [1]\\
  86. & $189.1,\ +3.0$ & $ 1.5-2.0$ & [1]\\
  87. & $190.9,\ -2.2$ & $1.0 $ & [1]\\
  88. & $205.5,\ +0.5$ & $1.6 $ & [1]\\
  89. & $206.9,\ +2.3$ & $2.2 $ & [1]\\
  90. & $260.4,\ -3.4$ & $2.2$ & [1]\\
  91. & $261.9,\ +5.5$ & $2.9$ & [1]\\
  92. & $263.9,\ -3.3$ & $0.25$ & [1]\\
  93. & $266.2,\ -1.2$ & $ <1$ & [1]\\
  94. & $80.2,\ -1.0 $ & $1.5 $ & [2]\\
  95. & $107.5,\ -1.5 $ & $1.1$ & [2]\\
  96. & $162.8,\ -16.0$ & $0.5 $ & [2]\\
  97. & $276.5,\ +19.0$ & $0.06-0.3$ & [2]\\
  98. & $287.4,\ -0.6$ & $ 3$ & [1]\\
  99. & $296.5,\ +10.0$ & $1.3-3.9$ & [1]\\
  100. & $308.8,\ -0.1$ & $2 $ & [1]\\
  101. & $309.8,\ +0.0 $ & $5$ & [1]\\
  102. & $315.4,\ -2.3$ & $2.3 $ & [1]\\
  103. & $317.3,\ -0.2$ & $ 4.0$ & [1]\\
  104. & $326.3,\ -1.8$ & $ 3.4-5.8$ & [1]\\
  105. & $327.4,\ +0.4$ & $4.3-5.4$ & [1]\\
  106. & $327.6,\ +14.6$ & $ 2.2$ & [1]\\
  107. & $330.0,\ +15.0$ & $0.15-0.5$ & [1]\\
  108. & $332.4,\ -0.4$ & $3.1 $ & [1]\\
  109. & $332.5,\ -5.6$ & $ 3.0$ & [1]\\
  110. & $337.3,\ +1.0$ & $ 5$ & [1]\\
  111. & $319.9,\ -0.7 $ & $2.5 $ & [2]\\
  112. & $323.9,\ -0.00$ & $3.1 $ & [2]\\
  113. & $343.0,\ -6.0$ & $ 1.5$ & [1]\\
  114. & $347.3,\ -0.5$ & $1.3$ & [1]\\
  115. & $350.1,\ -0.3$ & $4.5$ & [1]\\
  116. & $355.6,\ -0.0$ & $3$ & [1]\\
  117. & $359.1,\ -0.5 $ & $5$ & [1]\\
  118. & $348.9,\ 0.4$ & $3.6 $ & [2]\\
  119. & $350.2,\ 0.8$ & $0.27-0.39$ & [2]\\
  120. & $359.2,\ 0.8$ & $ 5$ & [2]\\
  120. & $335.2,\ +0.1$ & $ 1.8$ & [1]\\
  121. & $336.7,\ +0.5$ & $ 3$ & [1]\\
  122. & $353.6,\ -0.7$ & $2.4-6.1 $ & [1]\\
  123. & $343.1,\ -0.7$ & $ 2$ & [1]\\
  124. & $354.1,\ +0.1$ & $ 1.5$ & [1]\\
 \hline
\end{tabular}\\
\end{minipage}
\label{tab:SNRs}
\end{table*}


\bsp	
\label{lastpage}
\end{document}